\documentclass[useAMS,usenatbib]{mn2e}
\usepackage{graphicx,color,epsfig}
\usepackage{color}

\newcommand{\be}{\begin{equation}}
\newcommand{\ee}{\end{equation}}

\newcommand{\apj}{ApJ}

\newcommand{\mnras}{MNRAS}
\newcommand{\aap}{A\&A}
\newcommand{\araa}{ARA\&A}
\newcommand{\apjl}{ApJL}
\newcommand{\aj}{AJ}
\newcommand{\nat}{Nature}

\def\ltsima{$\; \buildrel < \over \sim \;$}
\def\simlt{\lower.5ex\hbox{\ltsima}}
\def\gtsima{$\; \buildrel > \over \sim \;$}
\def\simgt{\lower.5ex\hbox{\gtsima}}

\def\msun{{\,{\rm M}_\odot}}

\newcommand\mbh{{\,{\rm M}_{\rm bh}}}

\def\del#1{{}}

\title[Momentum feedback by black hole winds]{Simulations of momentum feedback by black hole winds}

\author[S. Nayakshin, C. Power]{Sergei Nayakshin$^{1,2}$ and
Chris Power$^1$\\ $^1$ Department of Physics \& Astronomy,
  University of Leicester, Leicester, LE1 7RH, UK \\
$^2$ Max-Planck-Institut  f\"ur Astrophysik,
Karl-Schwarzschild-Strasse 1, D-85741, Garching, Germany}

\begin{document}

\date{Accepted 2008 ?? ??. Received 2008 ?? ??; in original form 2008 05 ??}

\pagerange{\pageref{firstpage}--\pageref{lastpage}} \pubyear{2008}

\maketitle

\label{firstpage}

\begin{abstract} 
The observed super-massive black hole (SMBH) mass -- galaxy velocity
dispersion ($M_{\rm bh} - \sigma$) correlation may be established when
winds/outflows from the SMBH drive gas out of the potential wells of classical
bulges. Here we present numerical simulations of this process in a static
isothermal potential. Simple spherically symmetric models of SMBH feedback at
the Eddington luminosity can successfully explain the $M_{\rm bh} - \sigma$
and nuclear cluster mass $M_{\rm NC}-\sigma$ correlations, as well as why
larger bulges host SMBHs while smaller ones host nuclear star
clusters. However these models do not specify how SMBHs feed on infalling
gas whilst simultaneously producing feedback that drives gas out of the
galaxy.

More complex models with rotation and/or anisotropic feedback allow SMBHs to
feed via a disc or regions not exposed to SMBH winds, but in these more
realistic cases it is not clear why a robust $M_{\rm bh} - \sigma$ relation
should be established.  In fact, some of the model predictions contradict
observations. For example, an isotropic SMBH wind impacting on a disc (rather
than a shell) of aspect ratio $H/R \ll 1$ requires the SMBH mass to be larger
by a factor $\sim R/H$, which is opposite to what is observed. We conclude
that understanding how a SMBH feeds is as important a piece of the puzzle as
understanding how its feedback affects its host galaxy.

Finally, we note that in aspherical cases the SMBH outflows induce
differential motions in the bulge. This may pump turbulence that is known to
hinder star formation in star forming regions. SMBH feedback thus may not only
drive gas out of the bulge but also reduce the fraction of gas turned into
stars.

\end{abstract}


\section{Introduction}\label{intro}

It is believed that the centres of most galaxies contain super-massive black
holes (SMBHs) whose mass $\mbh$ correlates with the velocity dispersion $\sigma$ of
the host galaxy \citep{Ferrarese00,Gebhardt00,Tremaine02}. Similarly, there is
a correlation between $\mbh$ and the mass of the bulge,
$M_{\rm bulge}$ for large SMBH masses \citep{Magorrian98,Haering04,GultekinEtal09}. 
Observations also suggest that the masses of nuclear star clusters (NC) ($10^5 \msun
\simlt M_{\rm NC} \simlt 10^8 \msun$) correlate with the properties of their
host dwarf ellipticals \citep{FerrareseEtal06,Wehner06} in a manner that is
analogous to the one between SMBHs and their host ellipticals.

These empirical relations can be explained in a very natural way if 
the growth of host galaxies and their central SMBHs or NCs are linked
by feedback. This was first pointed out by \cite{SilkRees98}, who highlighted
the potential importance of SMBH heating and outflows before any robust observational 
evidence for such a link had been found. Subsequently, \cite{Fabian99} argued that radiation
pressure acting on cold gaseous
clouds in the bulge could give rise to the observed correlations. \cite{KP03}
argued for the existence sub-relativistic outflows from the very central
regions of AGN, which prompted \cite{King03,King05} to study the motion of a 
shell of gas swept up by a wind/outflow from a central black hole in a
galactic isothermal dark matter potential. \cite{King05} demonstrated that the
shell will be expelled from the potential provided the black hole mass exceeds
a critical value that, as a function of $\sigma$, turns out to be close to the 
observed $M_{\rm BH}$--$\sigma$ relation.

The main result of \cite{King03,King05} can be deduced using a simpler order
of magnitude ``weight argument''. According to this argument, the SMBH
luminosity is assumed to be limited by the Eddington value. Radiation pressure
drives a wind. \cite{KP03} argue that wind velocity is comparable to the
escape velocity from the inner accretion disc, e.g., $v \sim 0.1 c$. The
momentum outflow rate is assumed to be
\begin{equation}
\dot P_{\rm SMBH} \approx {L_{\rm Edd}\over c} = \frac{ 4 \pi G M_{\rm BH}}{\kappa}\;;
\label{pismbh}
\end{equation}
\noindent here $\kappa$ is the electron scattering opacity and $M_{\rm BH}$ is
the SMBH mass. This result is natural to the order of magnitude: $L_{\rm Edd}/
c$ is the radiation momentum flux which is presumably passed to the wind as
radiation accelerates the outflow \citep[cf.][]{KP03}.  The argument also
assumes that the black hole wind is optically thin at the point of interaction
with the ambient gas. Because the cooling time of the shocked gas is short on
scales appropriate for observed bulges \citep{King03,King05}, the bulk energy
of the outflow is thermalised and quickly radiated away. It is then only the
momentum push (equation \ref{pismbh}) of the outflow on the ambient gas that
is important since it is this that produces the outward force on the
gas\citep[as in the earlier model by][]{Fabian99}. The weight of the gas is
$W(R) = GM(R)[M_{\rm total}(R)]/ R^2$, where $M_{\rm gas}(R)$ is the enclosed
gas mass at radius $R$ and $M_{\rm total}(R)$ is the total enclosed mass
including dark matter. For an isothermal potential, $M_{\rm gas}(R)$ and
$M_{\rm total}(R)$ are proportional to $R$, so the result is
\begin{equation}
W = {4f_g\sigma^4\over G}\;.
\label{w}
\end{equation}
Here $f_g$ is the baryonic fraction and $\sigma^2 = GM_{\rm total}(R)/2R$ is
the velocity dispersion in the bulge.
By requiring that momentum output produced by the black hole just balances
the weight of the gas, it follows that
\begin{equation}
M_{\sigma} = {f_g\kappa\over \pi G^2}\sigma^4,
\label{msigma}
\end{equation}
\noindent which is consistent with the observed the $M_{\rm BH}$--$\sigma$ relation.

To order of magnitude, relation \ref{w} should hold for any potential at the
virial radius. Therefore the model by \cite{King03,King05} appears to be a
promising explanation of the observed correlations between SMBHs and their
host galaxies.  However, additional complications, not considered in
\cite{King03,King05}, may be important. Amongst these are (1) the self-gravity
of the gas, because gas that accumulates at the centre of the potential may
begin to dominate it and therefore alter the result; (2) finite angular
momentum of the gas may lead to formation of a disc; (3) collimated and
variable outflows. \\

The goal of this paper is to verify that the predictions of the analytical model of
\cite{King03,King05} hold and to explore more realistic settings, of the kind
just described. Note that we concentrate on the early stages of galaxy
evolution, when we would expect galaxies to be gas-rich and the baryonic mass
within dark matter haloes is dominated by gas rather than stars. At later
times, when the SMBH is likely to be less luminous and the galaxy is less
gas-rich, we would expect different forms of feedback to become important,
such as relativistic jets and the associated radio bubbles
\citep[e.g.,][]{ChurazovEtal02} or pre-heating due to the inverse Compton
effect \citep[e.g.,][]{SazonovEtal04}. However, we do not include these forms
of feedback in our current simulations.

\section{Fixed BH mass runs}\label{sec:fixed_mass}

\subsection{Numerical method}\label{sec:method}

\cite{NayakshinEtal09a} developed a new method for radiation transfer in SPH
based on Monte Carlo packets, which can also be used
to simulate winds in the momentum-conserving phase. As discussed in the
Introduction, one can consider only the momentum transfer from the outflow to
the ambient gas in this regime. In addition, for the problem at hand, the
typical velocity of the ambient gas is of the order the velocity dispersion in
the bulge, which is less than a few $100$ km/sec. In contrast, the black hole outflow is
much faster, $v \sim 0.1 c$, which implies that the mass of the black
hole wind is negligibly small compared with that of the gas in the bulge at
the point there the latter is driven away. Hence we can neglect the mass of
the black hole wind in comparison to that of the ambient medium.

Accordingly, we use massless particles moving with velocity $v=0.1 c$ to
simulate the black hole wind. The wind particles propagate in straight lines
until they encounter one or more SPH particles at which point they transfer
their momentum to the particles. For more information on the method
and validation of the code see \cite{NayakshinEtal09a}, especially
their \S 3.

For spherically symmetric tests, we use 35,000 SPH particles. The wind
particles carry momentum $p_{\rm wind} = 0.1 m_{\rm sph} \sigma$
each. This satisfies the requirement $p_{\rm wind} \ll p_{\rm sph}$,
where $p_{\rm sph}$ is the typical SPH particle momentum ($\sim m_{\rm
  sph} \sigma$ here), and ensures that Poisson noise from our Monte
Carlo scheme is small enough not to compromise our results
\citep[see][]{NayakshinEtal09a}. The particles are emitted by the
black hole at the rate
\begin{equation}
\dot N_{\rm wind} = \frac{L_{\rm Edd}}{c p_{\rm wind}}\;,
\label{nrate}
\end{equation}
where $L_{\rm Edd}$ is the Eddington luminosity for the black hole. They are
ejected isotropically unless stated otherwise.\\

We perform our simulations in a static singular isothermal sphere
potential with velocity dispersion $\sigma = 147$ km s$^{-1}$. The
units of mass and distance are $M_U = 10^{10}\msun$ and $R_U=1$ kpc
respectively; the unit velocity is $V_U = \sqrt{GM_U/R_U} = \sqrt{2}\sigma$. 
Equation \ref{msigma} yields the expected value for the critical mass at which
gas should be expelled;
\begin{equation}
M_{\sigma} = {f_g\kappa\over \pi G^2}\sigma^4 = 1.1\times 10^8 \msun\;.
\label{msigma0}
\end{equation}
To check this prediction, we first limit our simulations to spherically
symmetric initial conditions. We work with a finite extent shell since that
allows us to explore interesting regions of parameter space. The initial
radial distribution of gas follows the distribution of dark matter, assumed to
dominate the potential: $\rho(R) \propto R^{-2}$ for $R_{\rm in} < R < R_{\rm
  out}$. For all the tests in this paper, $R_{ \rm in} = R_{\rm out}/2$. As in
\cite{King03,King05}, the normalisation of gas density, and hence the mass of
the shell, is set by the cosmological baryon mass fraction, $f_{g0} =
\rho_g/\rho_{\rm total} \simeq 0.16$, where $\rho_{\rm total} = \rho_{\rm DM}
+ \rho_g$ is the total mass density, and $\rho_g$ and $\rho_{\rm DM}$ are the
gas and dark matter densities, respectively. For the isothermal potential,
$\rho_{\rm total}(R) = \sigma^2 /2\pi G R^2$. Because our focus is momentum
feedback, we adopt the isothermal equation of state for the gas, assuming $T =
3\times 10^5$ K. This is lower than the virial temperature, $T_{ \rm vir}
\approx 10^6$ K, for the potential we are using, preventing small scale gas
fragmentation. The latter process would lead to star formation and feedback
from star formation, in both heating and momentum-driven forms. We plan to
include these processes in our future work, for now neglecting them in
comparison with the SMBH feedback altogether.

We set $R_{\rm acc} = 0.5$ as the inner boundary of our computational
domain. SPH particles that cross inside this inner boundary are removed.
Note however that this scale is
still too large to associate it with SMBH accretion directly, and there is
no model-independent parameter-free way of connecting this ``sunken'' mass
with SMBH accretion and feedback. For simplicity we take the pragmatic
approach in this paper and consider only tests in which the black hole
feedback is either constant or increases at a rate fixed by the Eddington
limit. This allows us to concentrate on gas dynamics without worrying about
the complexities of the non-linear and model-dependent accretion-feedback
link. The ``accreted'' gaseous mass is thus considered to be mainly used in 
star formation inside $R_{\rm acc}$ and the feedback from that is neglected.\\

\begin{figure*}
\centerline{\psfig{file=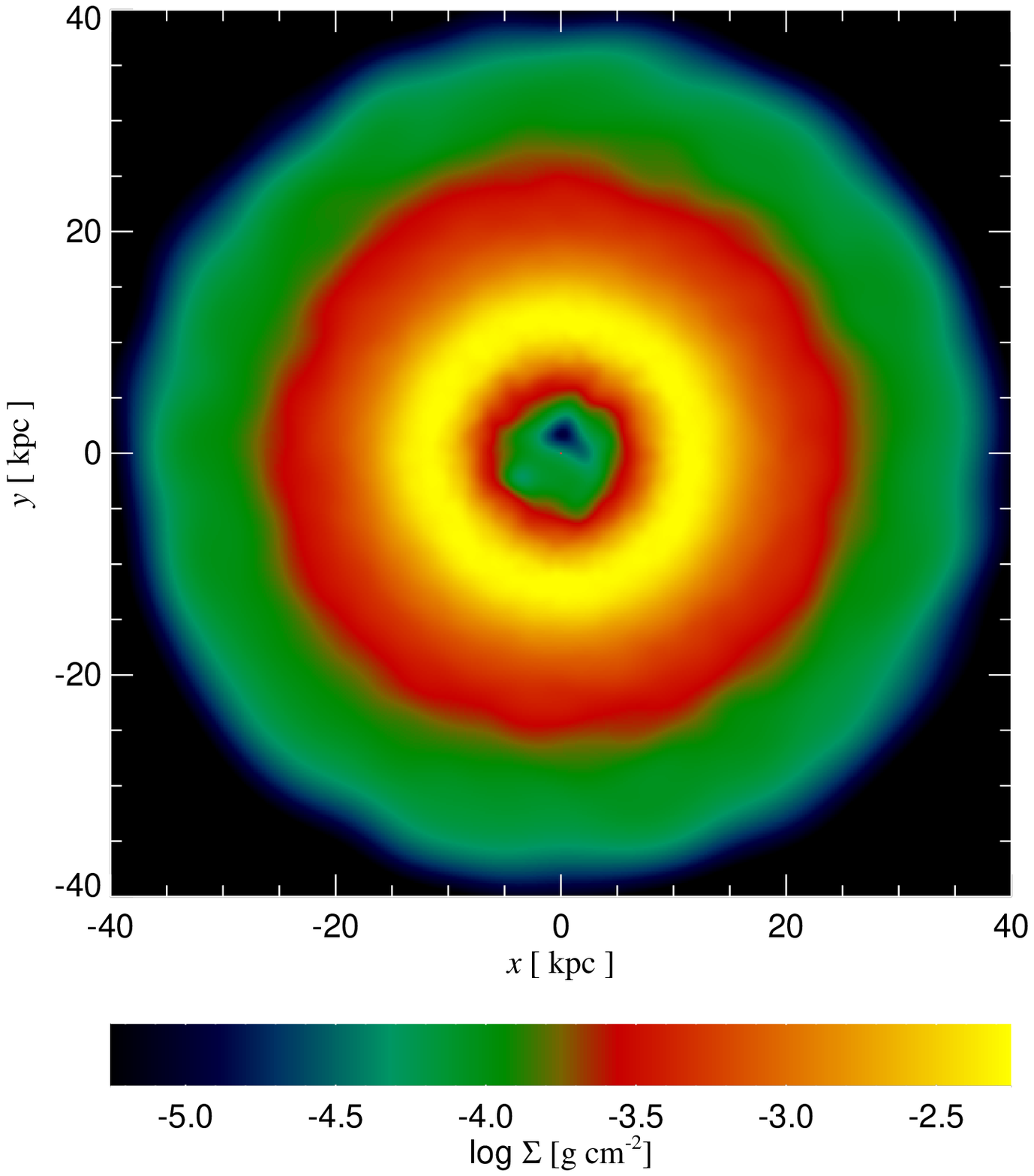,width=0.47\textwidth,angle=0}
  \psfig{file=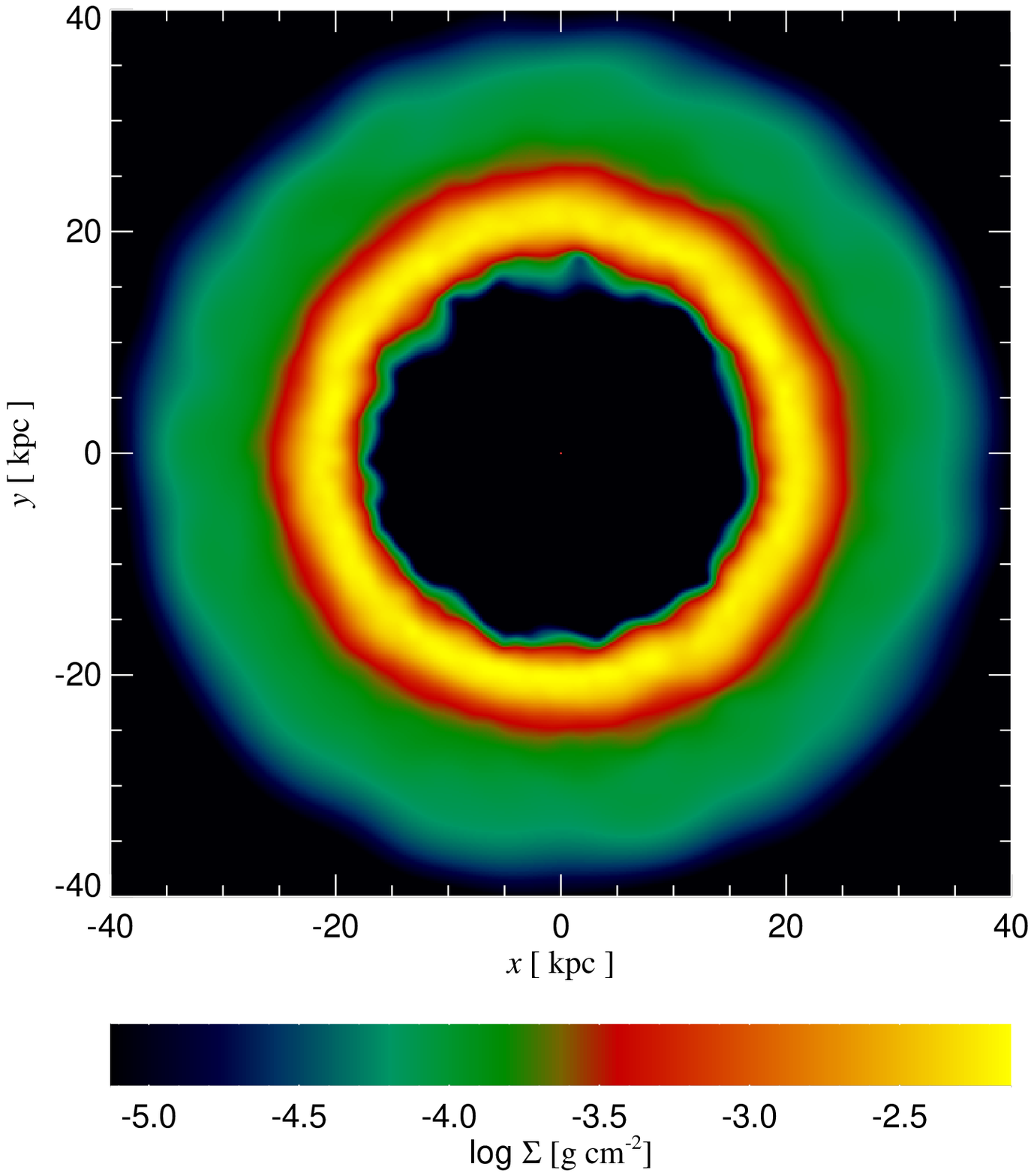,width=0.47\textwidth,angle=0}}
\caption{Typical results of the spherically symmetric tests. Left
  panel: shell collapsing on the $M_{\rm bh} = 2\times 10^8\msun$
  black hole (\S \ref{sec:underweight_fixed}). Right panel: same shell
  but the black hole mass is $M_{\rm bh} = 4 \times 10^8\msun$,
  sufficient to drive the shell out (\S
  \ref{sec:critical_fixed}). Both snapshots are done at time $t = 70$
  Myrs. See \S \ref{sec:method} for description of the figure.}
\label{fig:r40_fixed}
\end{figure*}

Before we begin the discussion of results, we show in Figure
\ref{fig:r40_fixed} two typical snapshots from the spherically symmetric
calculations. These show gas column densities in an angle-slice
projection. Specifically, the gas column density shown in the Figure
\ref{fig:r40_fixed}, and in all similar figures below, is calculated by
\begin{equation}
\Sigma(x, y) = \int_{-z(x,y)}^{z(x,y)} \rho(x, y) dz\;,
\label{sigma_proj}
\end{equation}
where the limits of the integration are given by $z(x,y)= r \tan\xi$,
and $r = (x^2 + y^2)^{1/2}$. The angle $\xi$ is chosen to be
$\tan\xi=1/4$ for this paper. This projection method is convenient as
it permits an unobstructed view into the central regions where the
black hole resides. Had we used a constant thickness $z(x,y)=$ const
projection method, then either the innermost region would not have
been resolved sufficiently or an insufficient number of the SPH
particles in the outermost regions would have been sampled for a
statistically meaningful figure.

The tests shown in Figure \ref{fig:r40_fixed} are described in \S
\ref{sec:underweight_fixed} (left panel) and \S \ref{sec:critical_fixed}
(right panel). Briefly, the former has $M_{\rm bh} = 2\times 10^8\msun$,
whereas the latter has $M_{\rm bh} = 4 \times 10^8\msun$. In both cases the
gas is falling in with radial velocity $v_r = - 1$ (in code units)
initially. In the lighter SMBH case, the black hole momentum outflow is
insufficient to reverse the shell's infall and it engulfs the black hole. In
the right panel, on the other hand, the initially thick shell is first
compressed to a thinner shell by the opposing actions of the black hole
outflow and the inward inertia of the outer layers, and then expelled to
infinity.

\subsection{Initially static shell tests}\label{sec:static}

\begin{figure}
\centerline{\psfig{file=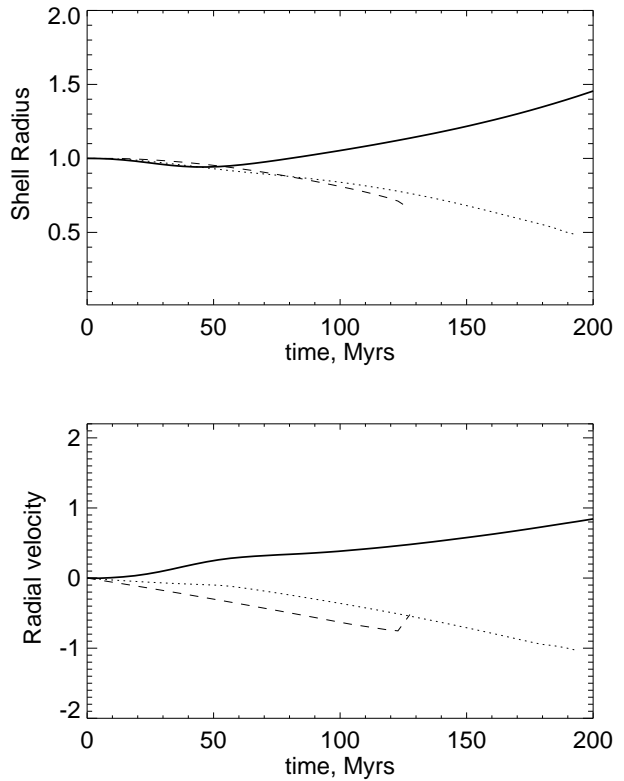,width=0.52\textwidth,angle=0}}
\caption{Mean radius (upper panel) and radial velocity (lower
  panel) of the gas as a function of time for three different black
  hole masses: $M_{\rm bh} = 2\times 10^8 \msun$, solid,$M_{\rm bh} =
  10^8 \msun$, dotted, and $M_{\rm bh} = 5\times 10^7 \msun$,
  dashed. The radius is plotted in units of the initial mean
  radius, whereas velocity is shown in code units ($\sqrt{2}
  \sigma$). See \S \ref{sec:static} for more detail.}
\label{fig:r_vs_t}
\end{figure}

We begin with tests in which gas is initially at rest (even though this is
unrealistic for the chosen temperature). The initial outer radius of the shell
is $R_{\rm out} = 40$ kpc for these runs. Figure \ref{fig:r_vs_t} shows
results of three such tests, where we vary the black hole mass from $M_{\rm
  bh} = 5\times 10^7 \msun$ to $M_{\rm bh} = 2\times 10^8 \msun$ by factors of
2. We define a mean radius and velocity of the gas by averaging over all the
SPH particles in a simulation. Defined in this way, the ``shell'' radius is plotted
in units of the initial mean radius, whereas the velocity is shown in code
units. It is apparent from Figure \ref{fig:r_vs_t} that the lower mass black
holes ($M_{\rm bh} \leq 10^8 \msun$) are unable to expel the shell of gas,
whereas the more massive black hole ($M_{\rm bh} \leq 2\times10^8 \msun$) is
able to expel the shell. We therefore
estimate that the critical black hole mass for this potential is around $1.5
\times 10^8 \msun$, in a very good agreement with the estimate in equation
\ref{msigma0}.

\subsection{Initially infalling shell, $M_{\rm bh} = 2\times
  10^8\msun$}\label{sec:underweight_fixed} 

This simulation is identical to that presented in \S \ref{sec:static} with
$M_{\rm bh} = 2\times 10^8 \msun$, except the radial velocity of gas is now
non-zero, with $v_r = - 1$. The additional initial inertia of the shell
suggests that the critical $M_{\sigma}$ mass should be higher in this case,
which turns out to be true.

\begin{figure}
\centerline{\psfig{file=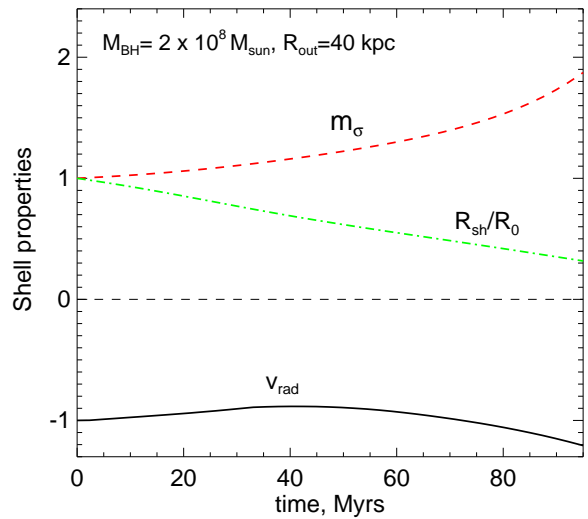,width=0.52\textwidth,angle=0}}
\caption{Mean radial velocity (solid) and radius (dash-dotted) of the gaseous shell as
  a function of time for $M_{\rm bh} = 2\times 10^8 \msun$. The dashed curve
  shows the expected dimensionless $m_\sigma$ mass, as defined by the equation
  \ref{sigmam_ratio}. The simulation is described in \S\ref{sec:underweight_fixed}.}
\label{fig:M0.02_R40}
\end{figure}

Figure \ref{fig:M0.02_R40} shows the mean shell radius, $R_{ \rm sh}$, in
units of its initial value (dash-dotted green curve) as a function of
time. The inward motion of the shell cannot be prevented in this simulation,
as is clearly indicated by the monotonically decreasing mean radius. The solid
curve shows the mean radial velocity of the gas scaled to its initial
value. The shell is slightly decelerated initially by the momentum of the SMBH
outflow, i.e., $|v_r|$ decreases until $t\sim 40$ Myrs. However, at $t\sim 40$
Myrs the shell starts to accelerate and falls ``onto'' the SMBH, completely
enveloping it. To follow the evolution of the shell after this point would
require much greater numerical resolution and, more importantly, the physics
of star formation.

The additional inward acceleration of the shell at $t\sim 40$ can be understood if one
considers the self-gravity of the gas. We calculate the velocity dispersion 
corrected for the shell's self-gravity:
\begin{equation}
\sigma_{\rm sh}^2(R) = \frac{G}{2R}\left[M_{\rm DM}(R) + M_{\rm sh}\right] = 
\left(1-f_{g0}\right) \sigma_0^2 + \frac{G M_{\rm sh}}{2R}\;,
\label{sigmashell}
\end{equation}
where $R=R_{\rm sh}$.  The second term increases for a constant $M_{\rm sh}$
and a decreasing $R_{ \rm sh}$. It is convenient to define a ``running'' value
of the expected $M_\sigma$ value based on the mean values of $\sigma_{\rm sh}$
and $R_{\rm sh}$ using equation \ref{msigma}. Further, we use the
dimensionless variable $m_\sigma$:
\begin{equation}
m_\sigma = \frac{M_\sigma(R_{\rm sh}, \sigma_{\rm sh})}{M_{\sigma 0}} =
\frac{f_g}{f_{g0}}\;\left( \frac{\sigma_{\rm sh}}{\sigma_0}\right)^4\;,
\label{sigmam_ratio}
\end{equation}
where $M_{\sigma 0} = M_\sigma(R_0, \sigma_0)$ is the initial value of the
$M_\sigma$ mass as a function of the initial values of shell radius $R_0$ and
velocity dispersion $\sigma_0$, and $f_g$ and $f_g$ is the current mass
fraction of gas inside radius $R$. The function $m_\sigma$ is plotted in
Figure \ref{fig:M0.02_R40} with the dashed curve for the present
simulation. It becomes immediately apparent that the SMBH will find it harder
to retard the shell once the shell has fallen in sufficiently deep into the
potential and the gas starts to dominate the gravitational potential.  If SMBH
mass is below the critical mass then a runaway radial contraction of the shell
occurs in this simple model.

\subsection{Initially infalling shell, $M_{\rm bh} = 4\times
  10^8\msun$}\label{sec:critical_fixed} 

\begin{figure}
\centerline{\psfig{file=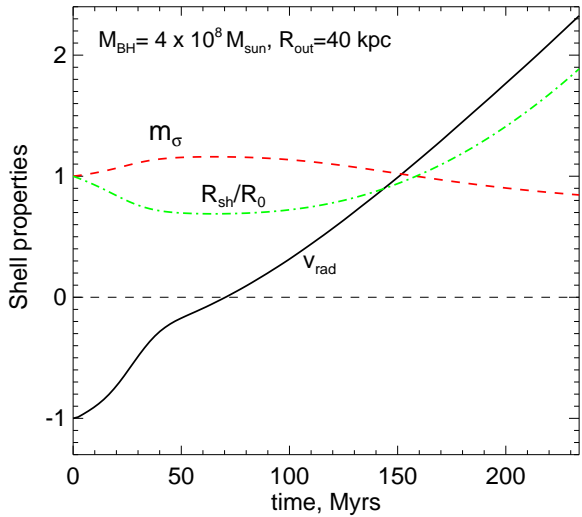,width=0.52\textwidth,angle=0}}
\caption{As in Figure \ref{fig:M0.02_R40} but now the black hole mass is a
  factor of two larger (\S \ref{sec:critical_fixed}). The gaseous shell is 
  now completely expelled from the potential well to infinity.}
\label{fig:M0.04_R40}
\end{figure}

Figure \ref{fig:M0.04_R40} shows the result of exactly the same calculation
but now with a SMBH mass twice as large, i.e., $M_{\rm bh} = 4\times 10^8
\msun$. The behaviour of the shell in this case is radically different. 
The average radius of the shell decreases only during the first 40 Myrs,
during which time the outermost layers of the shell continue to fall
in. However, at $t \sim 40$ Myrs the shell briefly decelerates before quickly
accelerating nearly linearly ($v_r \propto t$) out of the potential well.

It is worth noting how distinct the gas dynamics between the two cases
illustrated in Figures~\ref{fig:M0.02_R40} and ~\ref{fig:M0.04_R40}
respectively. This may be somewhat surprising given that the SMBH mass and
hence the outward acceleration differs between the cases only by a factor of
two. However, the key difference is that in the first case the shell fell in
far enough to become self-gravitating, with the self-gravity corrected
$\sigma$ increasing rapidly, increasing further the inward pull of
gravity. This effect, missing in the analytical theory of
\cite{King03,King05}, might actually increase the $M_\sigma$ value above that
given by equation \ref{msigma}.

Comparing the two initially infalling shell tests with those for an initially
static shell (\S \ref{sec:static}), we note that the critical black hole mass
is increased by a factor of two or so to $M_\sigma \approx 3\times 10^8
\msun$. This is natural as the shell needs to be accelerated to a positive
velocity $v_r \simgt 1$ to be driven out of the potential, and that is
comparable to the initial negative velocity of the infalling tests. The
required outward acceleration should then be roughly doubled compared with the
initially static tests. This shows that variations in the initial {\em radial}
velocity of the shell are unlikely to change the critical black hole mass by
more than a factor of a few.

\subsection{Scale-free nature of fixed BH mass
  solutions}\label{sec:scalefree} 

Equation \ref{msigma} shows that the critical $M_\sigma$ mass is independent
of the initial location of the shell, $R_0$, as long as the initial gas
fraction $f_{g0} = M_{\rm gas}(R_0)/M_{\rm tot}(R_0)$ does not vary with radius. We now
show that the same is true for time-dependent solutions. This can be seen
from the equation of motion for gas in the one zone approximation, which is
quite reasonable for a thin shell. We have
\begin{equation}
M_{\rm sh}\frac{d v_r}{dt} = \frac{L_{\rm Edd}}{c} - \frac{G}{R^2}\left[M_{\rm
      DM}(R) + M_{\rm sh}\right] M_{\rm sh}\;,
\label{dvrdt}
\end{equation}
where $v_r = dR/dt$. Let us assume $R(t) = R_0 g(\xi)$, where $\xi =
t/t_0$, $t_0 = R_0/\sigma$. Obviously, we require $g(0) = 1$. We can
now re-write the above equation in this form:
\begin{equation}
\frac{R_0}{t_0^2}\frac{d^2 g}{d\xi^2} = \frac{L_{\rm Edd}}{c M_{\rm sh}} -
\frac{2\sigma_0^2 (1-f_{\rm g0})}{R_0 g} - \frac{G M_{\rm sh}}{R_0^2 g^2}\;.
\label{dvrdt1}
\end{equation}
Only the dimensionless time variable $\xi$ remains, implying a scale-free solution.

\begin{figure}
\centerline{\psfig{file=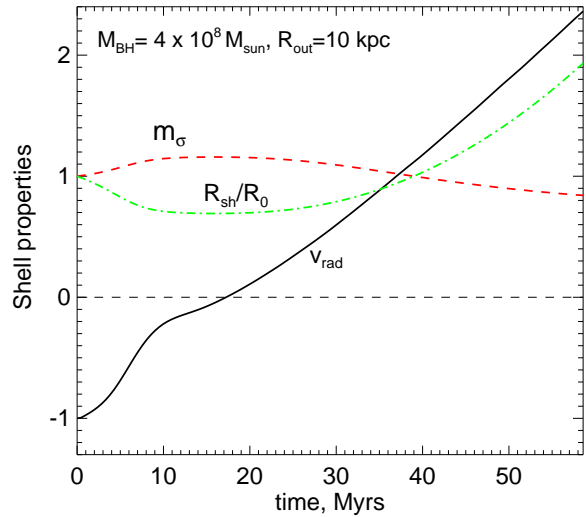,width=0.52\textwidth,angle=0}}
\caption{As in Figure \ref{fig:M0.04_R40}, but now the initial shell
  radius is 10 kpc rather than 40 kpc. Note that the result is identical
  to Figure \ref{fig:M0.04_R40} save for the time axis shrinking by factor of
  4. This confirms the scaling properties of the model discussed in \S
  \ref{sec:scalefree}.}
\label{fig:M0.04_R10}
\end{figure}

This conclusion is confirmed in Figure \ref{fig:M0.04_R10}, which shows the
result of a simulation identical to the one presented in Figure
\ref{fig:M0.04_R40}, except that the initial outer shell radius is $R_{\rm
  out} = 10$ kpc rather than $R_{\rm out} = 40$ kpc. The shell mass is also
reduced by a factor of four to keep $f_{g0}$ the same. It is readily apparent
that the two figures are identical, with the only difference evident in the
time coordinate.

This scale-free nature of the \cite{King03,King05} model in the isothermal
potential has interesting observational implications. The model predicts a
slope and normalisation for the $M_{\rm bh} - \sigma$ relation that is
consistent with the observed ones, assuming only one free parameter $f_{\rm
  g0}$ fixed at the cosmological baryon fraction; however the overall scale 
of the system needs to be set by additional physics. \cite{King03,King05}
points out that this can be the
requirement that the black hole outflow's cooling time is short compared with
the local dynamical time, which fixes the ratio of the black hole to bulge
mass to $\sim 10^{-3}$, in a good agreement with the observational data
\citep[e.g.,][]{Haering04}.

In principle, other factors can be affecting the black hole to bulge mass
relation, such as the finite extent of astrophysical potential wells, or the
amount of angular momentum in the gas. We shall now consider the first issue
in relation to the finite maximum growth rate of the black holes.

\section{Eddington rate -- limited growth models}\label{sec:eddington}

In \S \ref{sec:fixed_mass} the black hole mass was assumed fixed for
simplicity. In this section we allow the SMBH to grow at its Eddington
accretion rate. In this case the SMBH mass increases by the factor of $e$ in the Salpeter
time, 
\begin{equation}
t_{\rm Salp} = {M_{\rm bh} \over \dot M_{\rm Edd}} \approx 4.5 \times 10^7 \;\hbox{yrs}\;,
\end{equation}
where $\dot M_{\rm Edd} = L_{\rm Edd}/\epsilon c^2$ is the Eddington accretion
rate, and the radiative efficiency $\epsilon$ is set to 0.1.

The initial value of the outer radius, $R_{ \rm out}$, defines another important
time scale of the problem -- the dynamical time,
\begin{equation}
t_{\rm dyn} = \frac{R_{\rm out}}{V_U}\;,
\end{equation}
where $V_U = \sqrt{GM/R} \approx 208$ km/sec is the velocity unit for the
potential.  

As emphasised by \cite{NayakshinEtal09b}, there are two distinctly different
regimes. If $t_{\rm Salp} \ll t_{\rm dyn}$, then the SMBH {\em can} grow
arbitrarily quickly if provided with enough fuel. If there is a SMBH
feeding-feedback link that can limit the SMBH mass, such as in the
\cite{King03,King05} model, the latter then grows to the appropriate
$M_\sigma$ mass and remains there. If $t_{\rm Salp} \gg t_{\rm dyn}$, then
the SMBH is unable to grow sufficiently quickly to reach its maximum
(i.e. $M_\sigma$), even if it is provided with ample fuel during the dynamical
time of the system. We now present two simulations that explore these two
regimes.

\subsection{SMBH growth in a ``large'' bulge}\label{sec:large}

The initial condition used in this test is same as in \S
\ref{sec:underweight_fixed}, except that the initial SMBH mass is smaller,
$M_{\rm bh} = 10^8 \msun$.  As we found in \S \ref{sec:critical_fixed} for
this initial condition, the $M_\sigma$ mass is about $ M_0 = 3 \times 10^8
\msun$. The black hole thus needs to increase its mass by about a factor of 3
to drive the shell out. The dynamical time of the shell is $t_{\rm dyn}
\approx 200$ Myrs, which gives the SMBH plenty of time to grow. 

\begin{figure}
\centerline{\psfig{file=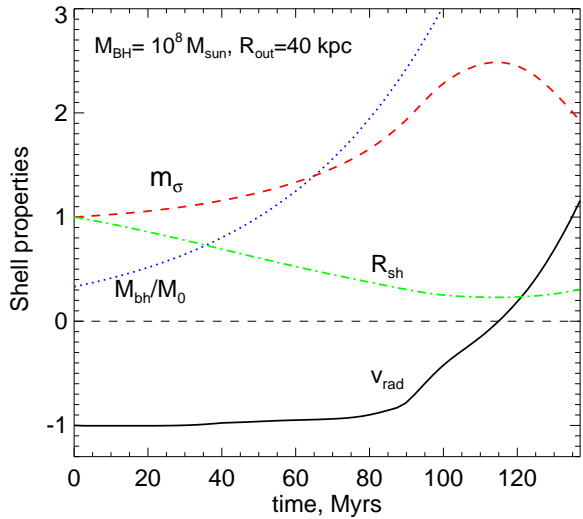,width=0.52\textwidth,angle=0}}
\caption{Gas mean velocity (solid), scaled mean radius (dot-dashed), scaled
  $M_\sigma$ (dashed) and the ratio of the SMBH mass to the initial $M_\sigma
  = M_0 = 3 \times 10^8 \msun$ mass (dotted) for an Eddington limited
  simulation (\S \ref{sec:large}). The initial SMBH mass is $10^8\msun$, a
  third of the $M_\sigma$ mass. However, due to rapid accretion the SMBH
  catches up with the growing (contracting) bulge and expels the gas when
  $M_{\rm bh} \approx 1.3\times 10^9 \msun$.}
\label{fig:M0.01_R40_edd}
\end{figure}

Figure \ref{fig:M0.01_R40_edd} shows the mean radial velocity of gas (solid
curve), the mean radius of the shell (dash-dotted), the self-gravity corrected
velocity dispersion of the gas (dashed) and finally, the dotted curve shows
the ratio of the black hole mass to the initial $M_\sigma = M_0 $ mass. 

The early phase of gas dynamics is quite similar to the underweight fixed mass
case considered in \S \ref{sec:underweight_fixed}. The shell is contracting
and the gas velocity dispersion grows with time, increasing the $M_\sigma$
value (see the dashed curve in the figure). However, the SMBH grows even
faster. At around 90 Myrs the gas suddenly gets decelerated and then
accelerated to positive velocity.

While this is as expected based on simple analytical expectations
\citep{NayakshinEtal09b}, the SMBH mass at the time when gas velocity becomes
positive is about $1.3\times 10^9 \msun$, about a factor of 4 higher than the
initial configuration value, $M_\sigma = 3 \times 10^8 \msun$. At the same
time, the shell radius is much smaller at that moment than the initial value,
increasing the self-gravity corrected $M_\sigma$ value by a factor of about
2.5 to $\sim 7.5 \times 10^8 \msun$. If our models included star formation and
if a good fraction of gas was turned into stars then the resulting bulge,
assuming the stars remain bound as the gas is blown away, would satisfy the
$M_{\rm bh} - \sigma$ relation within a factor of two. However the bulge
velocity dispersion would be higher than that of the underlying isothermal
potential value.

\subsection{SMBH growth in a ``small'' bulge}\label{sec:small}

\begin{figure}
\centerline{\psfig{file=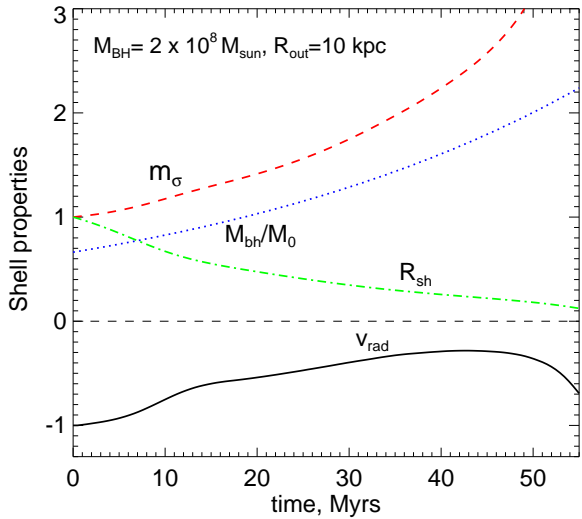,width=0.52\textwidth,angle=0}}
\caption{Similar to the simulation shown in Figure \ref{fig:M0.01_R40_edd} but
  with a smaller initial radius of $R_{\rm out} = 10$ kpc, and a larger
  initial SMBH mass of $2\times 10^8 \msun$. Note that the feedback from the
  growing SMBH is unable to prevent the contraction of the shell in this
  simulation.}
\label{fig:M0.02_R10_edd}
\end{figure}

We now repeat the run of \S \ref{sec:large} but shrinking the shell's outer
radius by a factor of four to $R_{\rm out} = 10$ kpc and increasing the SMBH's
initial mass to $M_{\rm bh} = 2\times 10^8 \msun$. This is about two thirds 
of the initial $M_\sigma$ mass, and hence the black hole needs to increase
in mass by only a small fraction to reverse the inflow of gas. However, as
Figure \ref{fig:M0.02_R10_edd} demonstrates, the black hole's growth is too
slow for this configuration of gas. As SMBH mass grows, so does the required
$m_\sigma$ mass, since the shell contracts. In fact, when the shell's mass
exceeds the local dark matter mass the shell becomes self-gravitating and the
increase in $m_\sigma$ accelerates, leaving no chance for the SMBH to catch
up. 

The results of these experiments confirm that the Salpeter time should be
sufficiently short compared to the the dynamical time of the system in order 
for the SMBH feedback to compensate for the shell contraction. However, if
star formation had been included, removing some of the available gas and
producing its own feedback, this requirement could be relaxed somewhat.

\section{Feedback on gas with angular momentum}\label{sec:rotation}

So far we have neglected the angular momentum of the gas, instead looking at
problems in which the motion of the shell is purely radial. However, we expect
the gas to have non-zero angular momentum and this will be as important for
the impact of the SMBH feedback, if not more so, than how the the gas is
distributed or what its thermodynamical properties are. If all the gas in the
bulge possesses non-zero angular momentum with respect to the black hole, then
its feeding must proceed through a disc. As efficiency of disc accretion is
not well understood in the presence of star formation in massive cold discs
\citep[e.g.,][]{Goodman03,NayakshinEtal07}, the exact distribution of gas in
the the angular momentum space prior to disc formation is very important.
Furthermore, on larger scales the angular momentum of gas determines the
resulting scale of the galactic disc, and perhaps the bulge.

The parameter space of this problem is very large, and so we can only begin to
scratch the surface of the problem here. We choose to study a case of the same initially
spherical shell infalling with $v_r = -1$ at $t=0$, $R_{\rm out} =40$ kpc
(e.g., \S \ref{sec:underweight_fixed}), but now rotating around the $z$-axis.
In the spherical coordinates, in which $z = r \cos \theta$, $x = r \sin
\theta\cos\phi$, $y = r \sin \theta\sin\phi$, the rotation law is given by
\begin{equation}
v_\phi = v_{\rm rot} \sin \theta\;,
\label{vphi}
\end{equation}
where $v_\phi$ is the $\phi$-component of velocity, and $v_{\rm rot}
=0.3$. Thus material near the pole rotates slowly while material near the
equator ($z=0$) rotates at the maximum for the test. We expect these tests to
be geometrically more complicated than the spherical shell tests. Therefore,
we significantly improve the resolution of these simulations by employing
$N_{\rm SPH}\approx 4\times 10^5$ particles, i.e., by more than a factor of
10.

\subsection{Gas dynamics without feedback}\label{sec:nofb}

\begin{figure*}
\centerline{\psfig{file=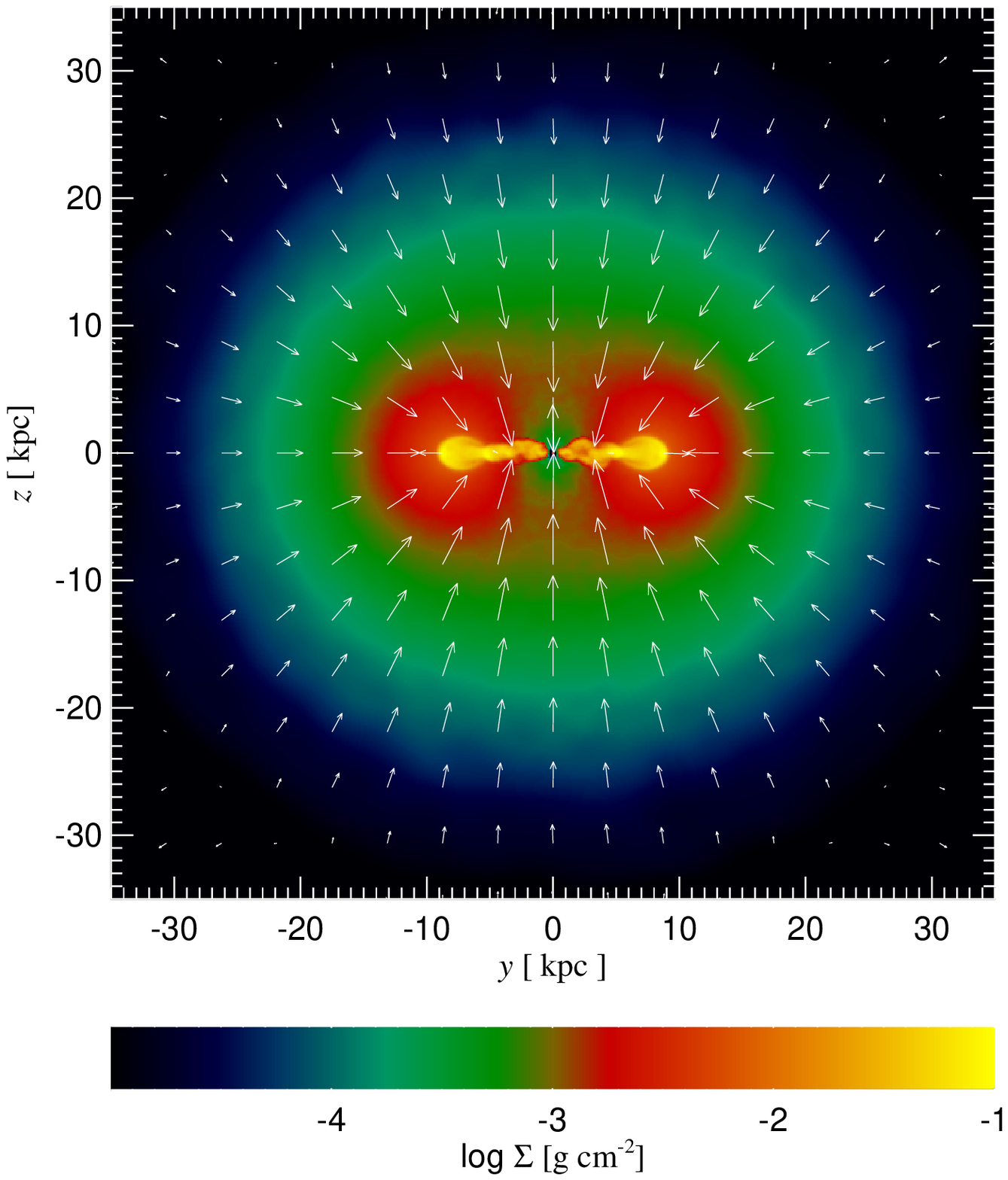,width=0.5\textwidth,angle=0}
\psfig{file=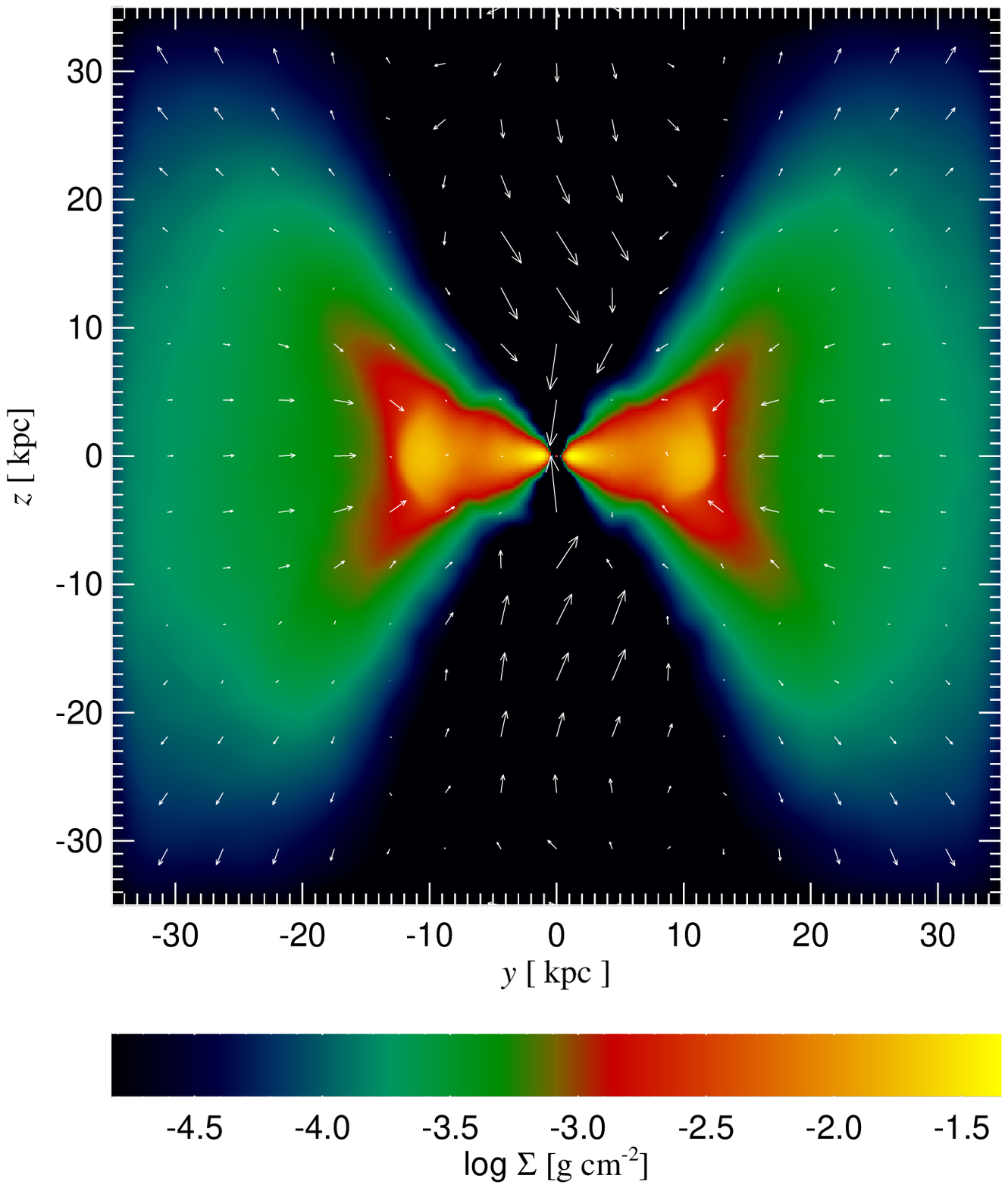,width=0.5\textwidth,angle=0}}
\caption{Column density in the edge-on projection of the collapsing
  shell for simulation without feedback. Left: time $t=95$
  Myrs. Right: $t=230$ Myrs. The final relaxed disc is hydrostatically
  supported at $H/R\sim 0.3$. Note the high density inner disc. About
  4\% of the shell has been accreted by the inner boundary at $R=0.5$ kpc.}
\label{fig:side_L0}
\end{figure*}

We begin by studying the dynamics of the gas in the absence of feedback from
the black hole. Figure \ref{fig:side_L0} shows two snapshots of the simulation viewed
edge-on. We also show velocity vectors of the gas. In this figure and in all
others showing velocity information, the normalisation of velocity vectors is
not to unit velocity but to the largest velocity found in the snapshot. This
is convenient as a subset of gas can have a velocity much larger than unity in
the more complex cases considered below, and its velocity vectors then clutter
the figures.

As expected, a gaseous disc forms with an outer radius of about $R \sim R_{\rm
  out}/3\sim 13$, appropriate for the rotation velocity equal to about a third
of the circular velocity in the isothermal potential. The inner radius of the
disc goes all the way to our inner boundary of the simulation domain, $R
=0.5$ kpc, since the regions close to the polar axis have small angular
momentum. By late times about 4\% of the gas disappears inside the inner
boundary. The disc undergoes radial oscillations for many orbits before the
gas settles on circular orbits. Signatures of these oscillations can be
observed in the right panel of Figure \ref{fig:side_L0}, where the disc
vertical structure is not completely relaxed. In particular, one can see some
low density regions having a small positive radial velocity in the corners of
the Figure.

Due to the high gas temperature selected for the simulations, $T = 3\times
10^5$K, the vertical pressure scale-height of the final disc is relatively
large, $H/R \sim 0.3$. Realistic discs are expected to be cooler, and thus
could be geometrically thinner and denser. However star formation in these
discs and especially feedback due to star formation may make such discs
vertically more extended than one would get from simple estimates based on
their temperature only. In any case, our deliberate choice of a rather
geometrically thick disc implies that our main conclusions on the role of
angular momentum, to be drawn below, can only be strengthened if discs are
thinner and denser.

\subsection{Feedback and disc formation}\label{sec:feedback_am}

We now turn to simulations with black hole winds. In order to isolate the main
effects in the interplay between the feedback and angular momentum, we
consider the simplest case of a constant and isotropic feedback. The momentum
flux from the black hole is fixed at the value appropriate for a black hole
mass of $M_{\rm bh} = 2 \times 10^8 \msun$, as in \S
\ref{sec:underweight_fixed}.

\begin{figure*}
\centerline{\psfig{file=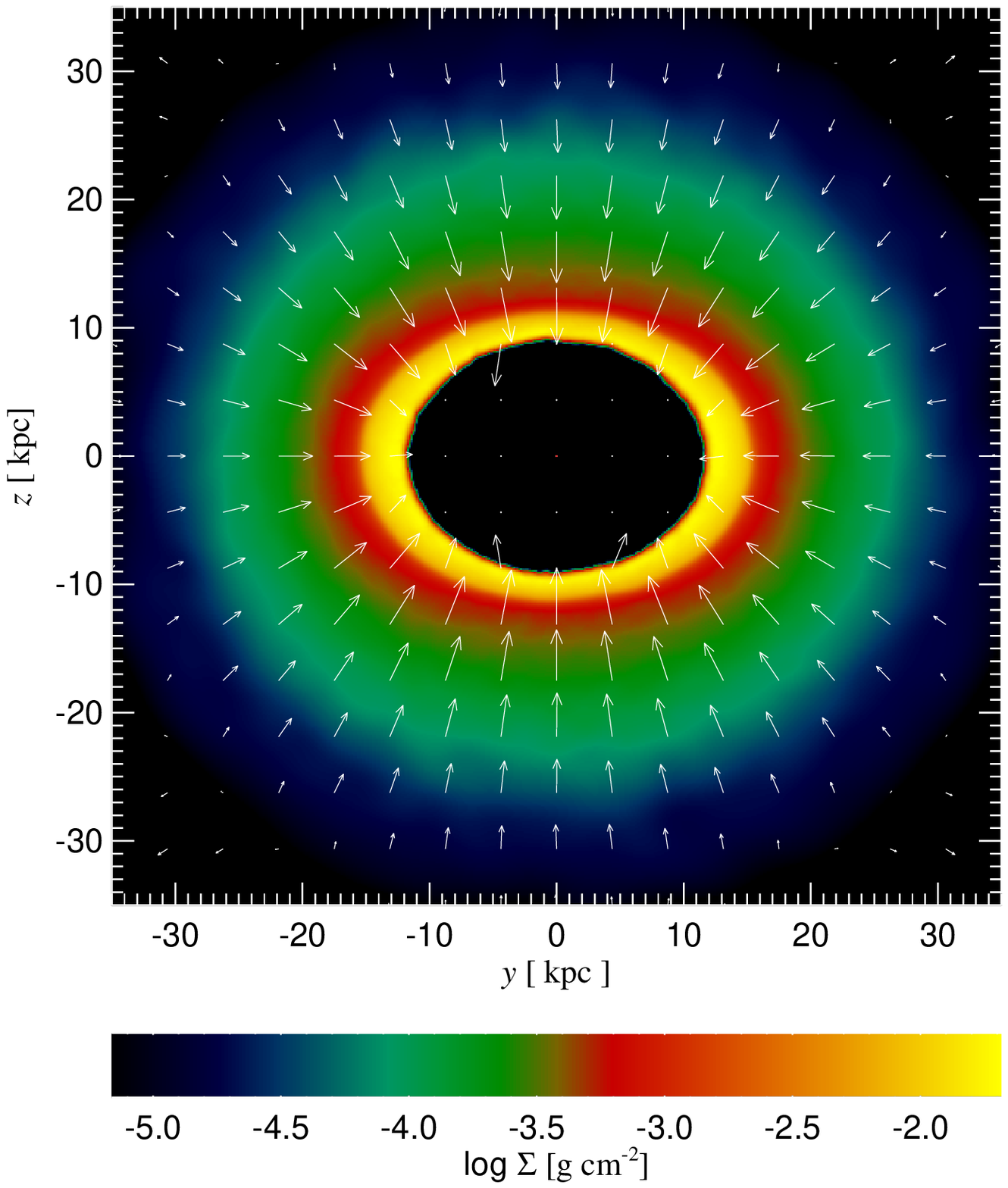,width=0.5\textwidth,angle=0}
\psfig{file=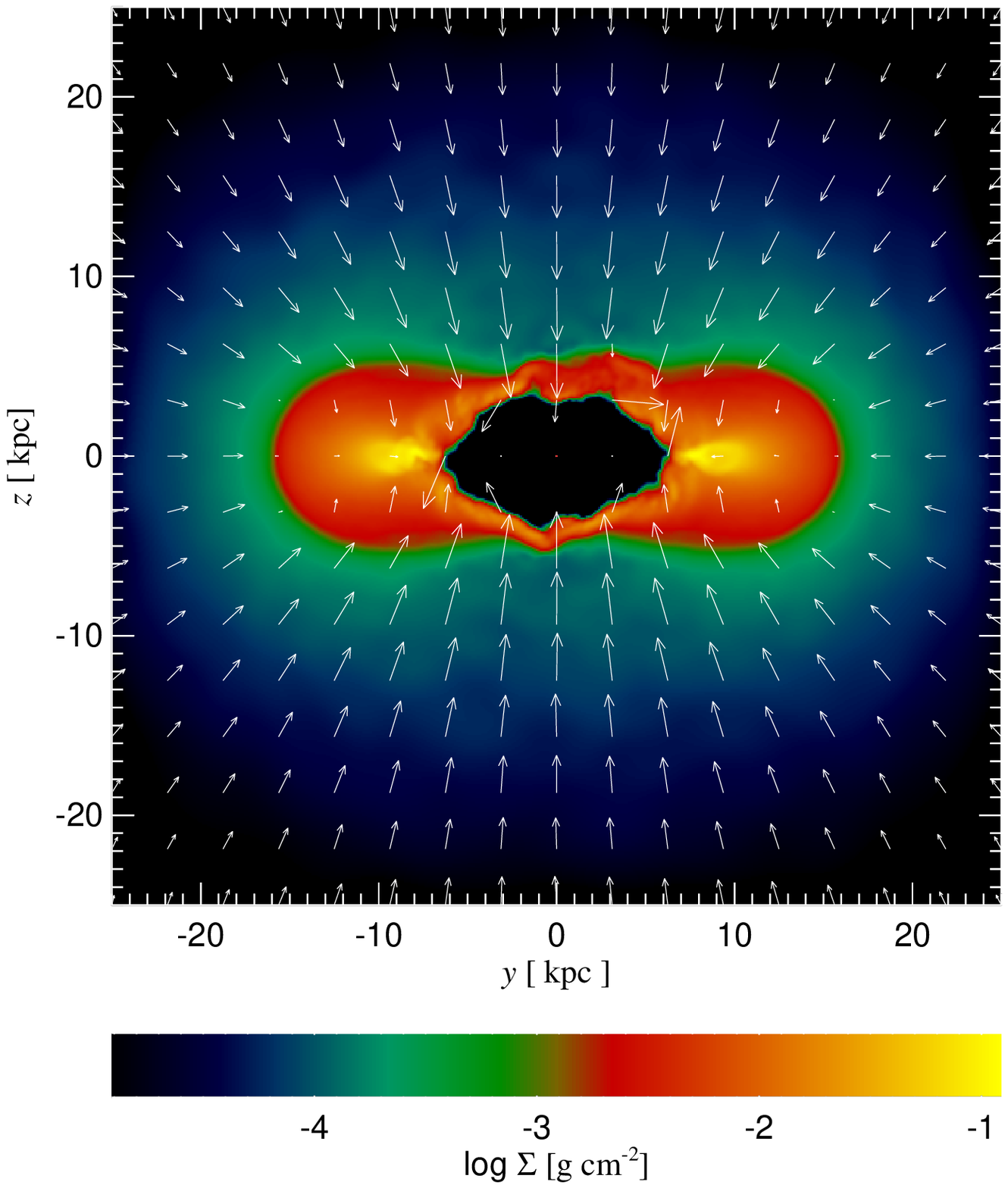,width=0.5\textwidth,angle=0}}
\caption{As in Figure \ref{fig:side_L0} but with feedback, and at
  times $t=95$ Myrs (left panel) and $t=130$ Myrs (right panel).}
\label{fig:side_L1}
\end{figure*}

Figure \ref{fig:side_L1} shows two snapshots from the simulation. The left
panel of Figure \ref{fig:side_L1} demonstrates a rather obvious point -- that
inflow is slower in the presence of feedback than in its absence (Figure
\ref{fig:side_L0}); in fact, the inner region is devoid of gas. As in the
spherically symmetric feedback cases, a shell is formed. The shell is not
spherical. The equatorial regions are held back by the centrifugal force more
than the polar regions. Whilst this is occurring, gas also starts to
accumulate in the midplane, just like it does when there is no
feedback. Because there is a competition between feedback and centrifugal
forces on the one hand, and gravity of the other, a ring -- rather than a disc
-- is seen to form just beyond the shock front in the right panel of Figure
\ref{fig:side_L1}. Also note that the cavity is not perfectly azimuthally
symmetric at later times (right panel), due to the development of a
combination of instabilities and the self-gravity of the gas.

\begin{figure*}
\centerline{\psfig{file=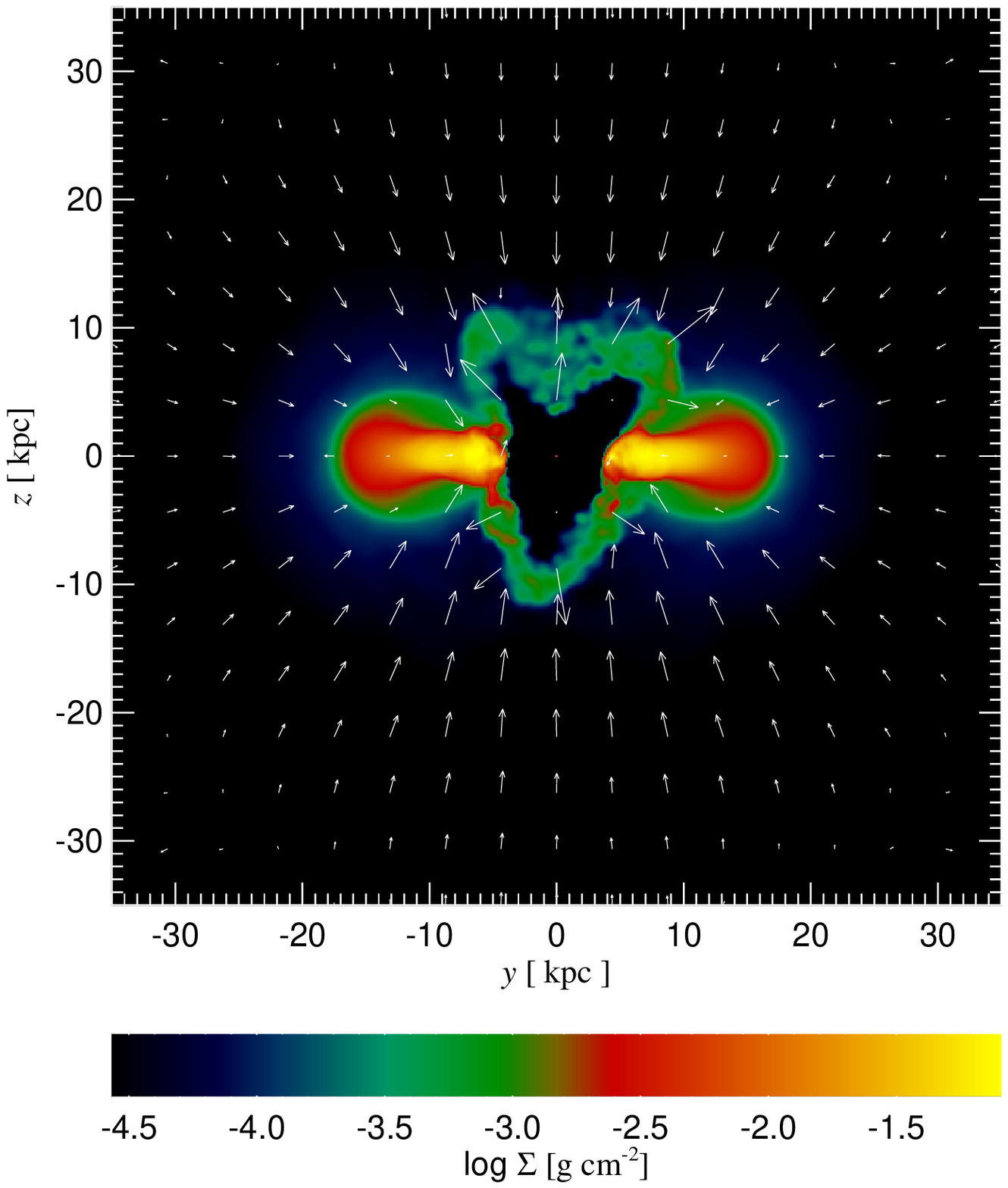,width=0.5\textwidth,angle=0}
\psfig{file=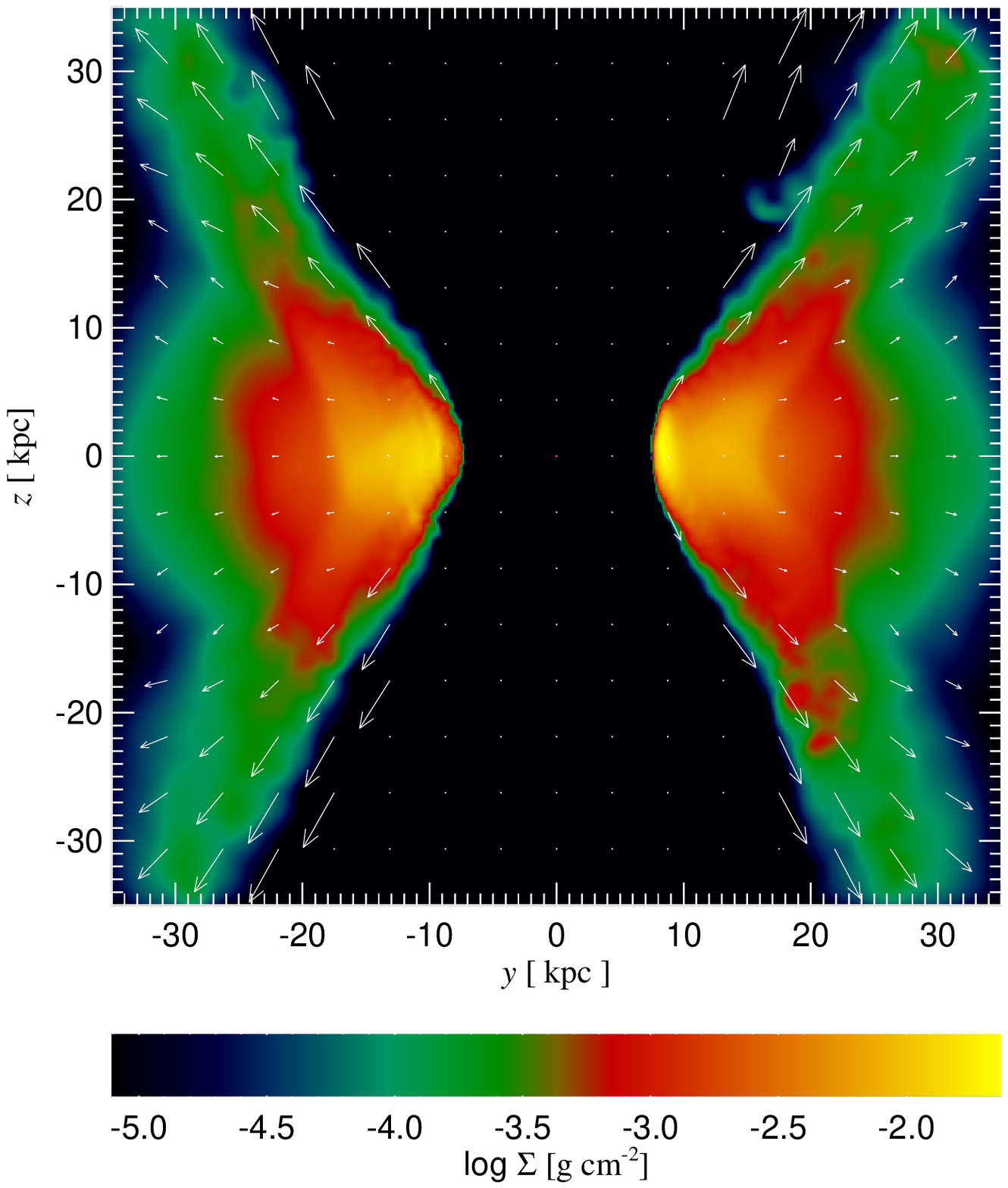,width=0.5\textwidth,angle=0}}
\caption{As in Figure \ref{fig:side_L1} but for later times. Left:
  time $t=145$ Myrs. Right: $t=275$ Myrs.}
\label{fig:side_L1_later}
\end{figure*}

Figure \ref{fig:side_L1_later} shows two snapshots at a later time.  A bipolar
initially bubble-like structure develops around but not exactly on the
symmetry axis. The bubble quickly bursts and a cavity is opened along
the symmetry axis. In that region some gas is driven sideways, by
the centrifugal force, and some to infinity by the black hole wind. At
later times the central region $R\simlt 10-15$ is devoid of gas except for the black hole
wind. Gas on the interface between the void with the higher density
gas is continuously stripped away, feeding the (secondary) outflow.

The centrifugal force rarefies the region of the shell near the poles, and the
black hole winds then evacuate that region by pushing the gas away. In the
equatorial plane, on the other hand, the disc becomes too dense for the black
hole feedback to have much of an effect. The disc manages to shadow its
interior from the black hole influence.

Analysing the net effect of the finite angular momentum of the shell on the
feedback--shell interaction, we note that the polar regions can now be
expelled more easily. If these regions are to ultimately provide the SMBH with
fuel then this would reduce the required $M_\sigma$ mass. However, it is now
much more difficult for the SMBH to affect the self-shielding disc. The mean
density in the disc is $\sim R/H$ times higher than the density for the same
mass spread in a spherical volume. If it is the disc that provides the SMBH
with fuel, then one would predict a value for the $M_\sigma$ mass higher than
the equation \ref{msigma} by at least the factor of $R/H$. Indeed, in case of
isotropic feedback, the fraction of SMBH wind intercepted by the disc is $\sim
H/R$. If feedback is anisotropic and beamed away from the disc, as would seem
natural for a ``grand design'' disc, i.e., one extending from galaxy scales of
kpc to sub-pc scales, then there is a further inefficiency in feedback
delivery to the disc. This appears to contradict the recent observations of
``pseudo bulges'', e.g. bulges that are disc or bar-like. The SMBH masses in
such bulges appear to be smaller rather than larger than those in classical
bulges at the same velocity dispersion \citep{Hu09}.

In fact it is hard to see why there would exist any $M_{\rm bh} - \sigma$
relation at all if SMBH were fed by large scale discs that are immune to the
SMBH feedback.

\section{Collimated feedback}\label{sec:jet}

There are good reasons to believe that feedback may be collimated rather than
isotropic. Unlike stars that radiate due to internal energy sources, black
holes need a continuous fuel supply in the form of accreting gas. Thus there
must be regions where feedback is relatively inefficient and net inflow
dominates. Most likely such inflow takes the form of a standard accretion disc
\citep{Shakura73} on scales of mpc and less. The outflow is then probably
stronger along the axis of symmetry of the disc. In addition, the putative
molecular torus is probably quite a massive and geometrically thick structure
\citep[e.g.,][]{Krolik88} that may restrict the black hole outflow to the
direction perpendicular to the midplane of the torus (if one can be
defined). Note that the inner accretion flow does not have to be co-aligned
with the torus, especially if the black hole is rapidly spinning
\citep[][]{KingEtal05}. If direction of the flow fluctuates rapidly compared
with the time scales of bulge formation, $\sim 10^8$ years, then we arrive at
possibly quite a complex picture of black hole outflows -- non-stationary,
non-spherical and with a fluctuating direction.

How is the momentum feedback picture developed for the spherical model
\citep{King03,King05} modified when these collimation effects are
taken into account? The answer is probably very complicated and
depends on the setting/environment. Here we present simulations for
two relatively simple situations, designed to motivate further discussion
rather than give final answers.

\subsection{Outflow misaligned with the
  disc}\label{sec:misalligned}

Our first simulation uses a fixed black hole mass $M_{\rm bh} =
2\times 10^8 \msun$ for simplicity, and the same rotating shell initial
condition used in the simulations presented in \S
\ref{sec:feedback_am}. However, rather than assuming isotropic feedback, 
the black hole feedback is now uniformly
distributed within the angle $0.7 \le |\cos\theta'| \le 1$ around the symmetry
axis $z'$ (see below). Note that
within this conical region the momentum flux density carried by the
black hole wind is $1/0.3\approx 3$ times higher than in the isotropic
case studied in \S \ref{sec:feedback_am}.

We anticipate that the effect of collimated feedback when the axis of symmetry
coincides with the $z$-axis should be qualitatively similar to the effect of
isotropic feedback on gas with non-zero angular momentum, of the kind
presented in (\S \ref{sec:feedback_am}). Therefore we consider a less trivial
case in which the direction of the axis of symmetry of the outflow, $z'$, is
inclined by angle $\pi/4$ from the $z$-axis defined by the direction of the
angular momentum vector of the shell. For convenience of presentation, we
choose the angular momentum vector of the shell and the outflow symmetry axis
to lie in the $z-y$ plane.

\begin{figure*}
\centerline{\psfig{file=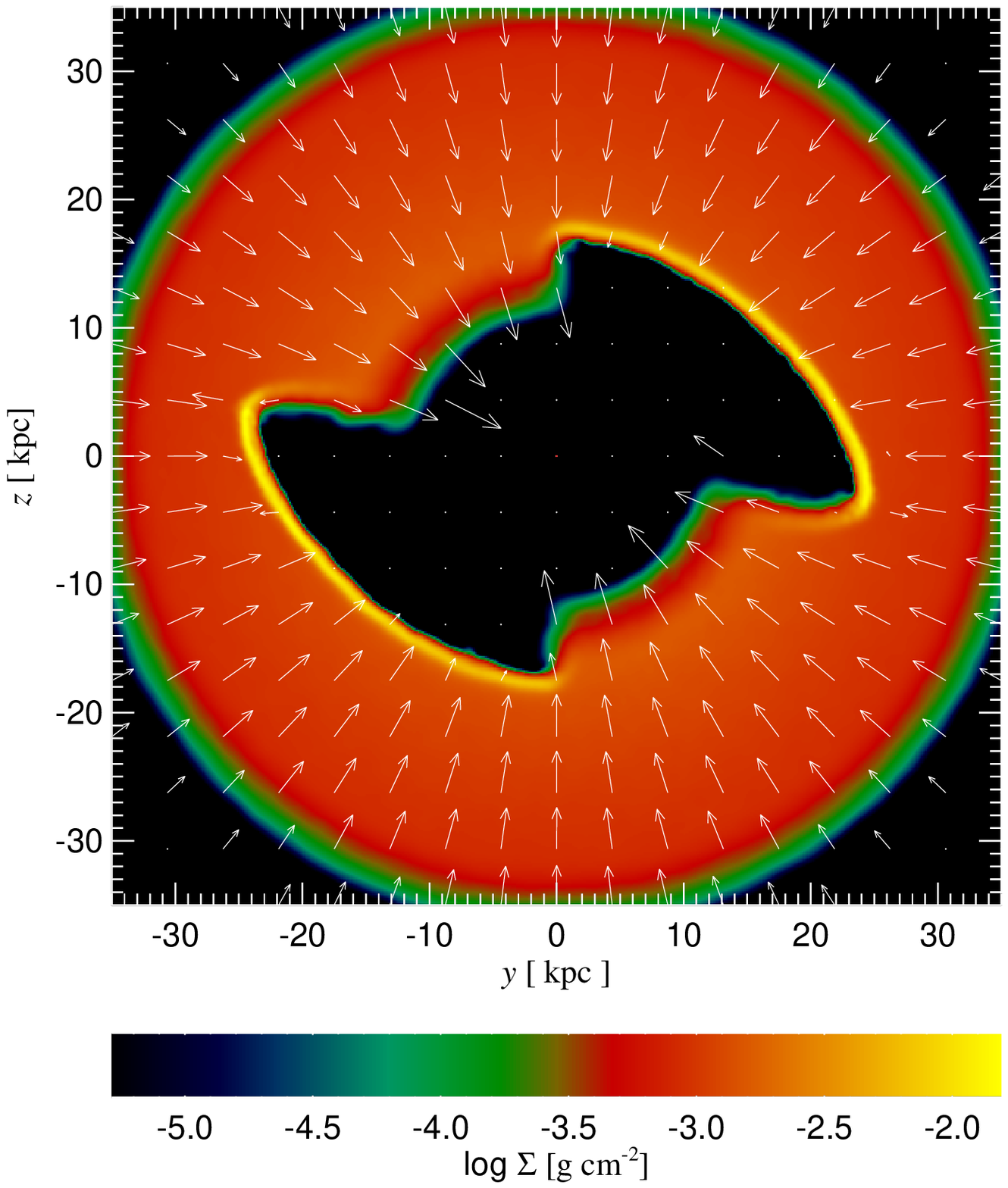,width=0.5\textwidth,angle=0}
\psfig{file=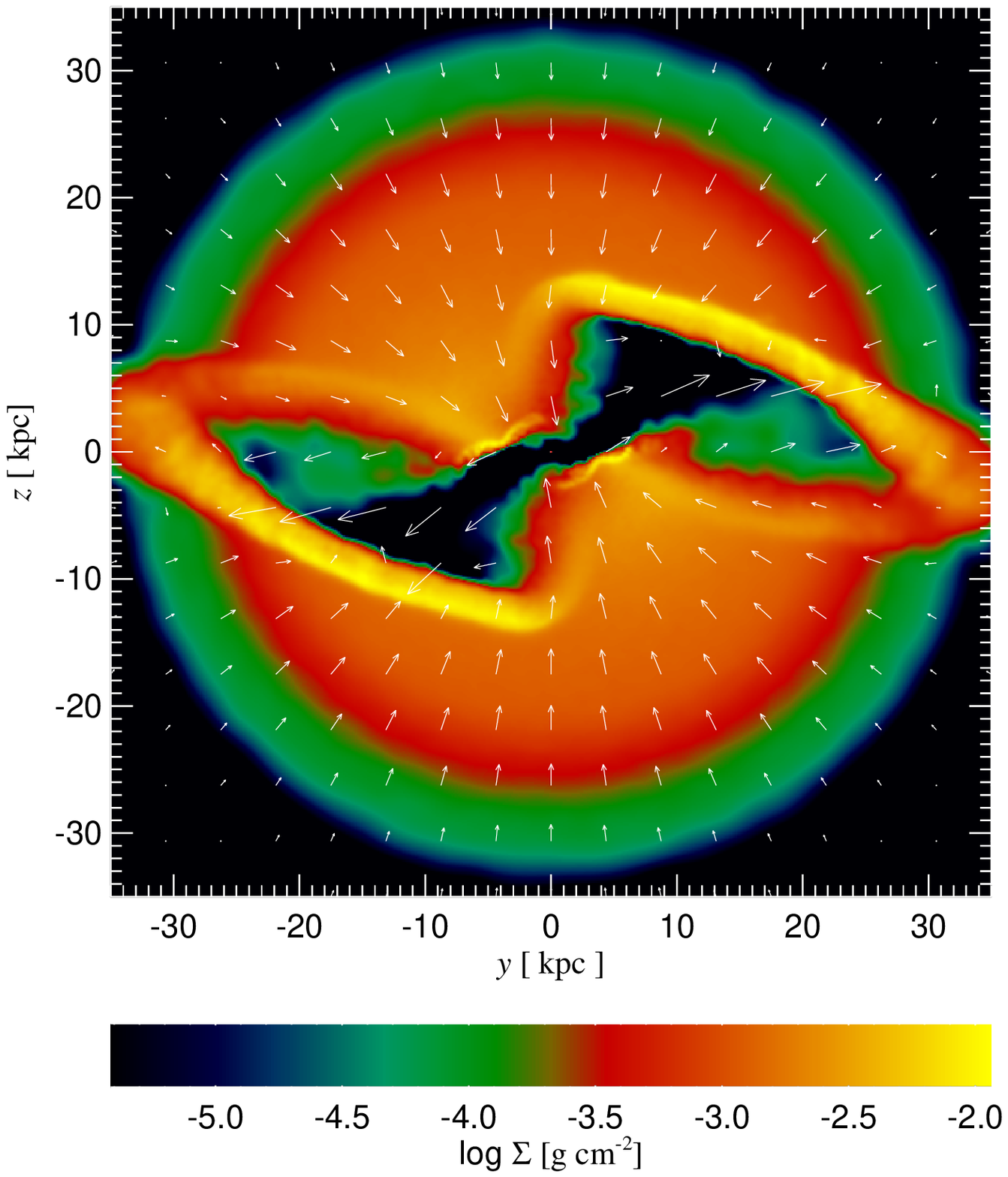,width=0.5\textwidth,angle=0}}
\caption{Angle-slice projected densities for the misaligned
  simulation at times $t=23$ (left panel) and $t=59$ (right panel)
  Myrs.}
\label{fig:angle45_fig1}
\end{figure*}

Figure \ref{fig:angle45_fig1} shows angle-slice projections of the gas surface
density at times $t=23$ Myrs (left panel) and $t=59$ Myrs (right panel). The
projections are done along the $x$-axis, as before. In this projection both
the disc axis of rotation ($z$-axis) and the outflow axis of symmetry are in
the plane of the figure. It is apparent from Figure \ref{fig:angle45_fig1}
that the outflow drives strong non-spherically symmetric motions in the
infalling shell, sending gas on different non-circular orbits inclined to the
$z$-axis at various angles. 

The evolution of the shell is different from anything seen in the previous
tests. Initially (see the right panel of the figure) regions of the shell
directly exposed to the feedback are compressed into a thin shell, which is
slowly driven outwards. However, because the shell is rotating, gas that was
not previously exposed to feedback continuously replenishes the outflow's
cone, and most of the shell eventually collapses inwards rather than being
driven outwards, as is seen in the right panel of Figure \ref{fig:angle45_fig1}.

At the same time, the segments of the shell not exposed to feedback at all --
regions along the diagonal line from the left upper corner to the right lower
corner of the figure -- approach the black hole initially almost freely
falling (the right panel). At later times their radial motion is stopped by
the centrifugal force. Some parts of the gas then appears to be pushed into
the feedback cone by the ram pressure of the continuous inflow. These parts are
blown away into the outflow.

\begin{figure*}
\centerline{\psfig{file=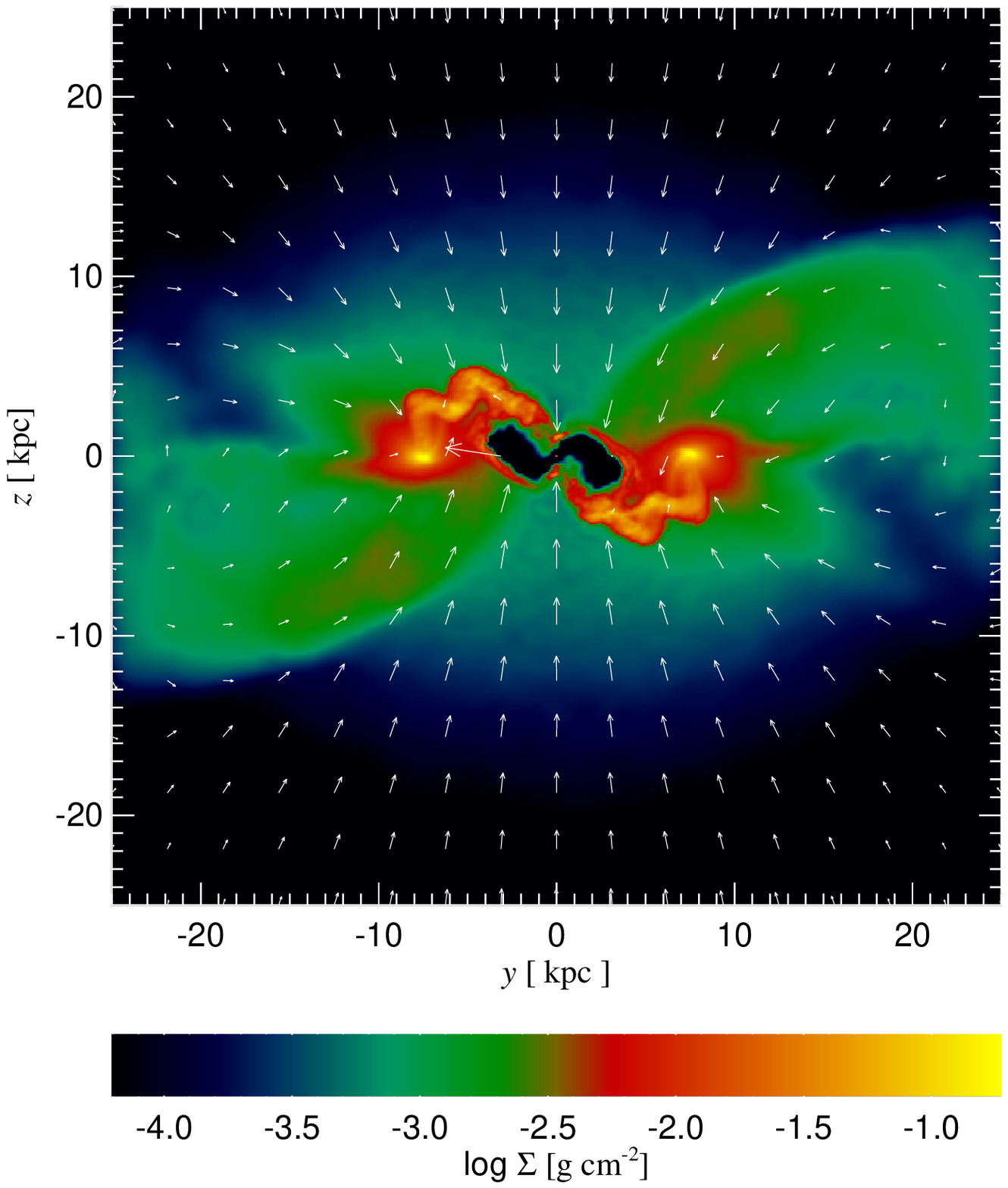,width=0.5\textwidth,angle=0}
\psfig{file=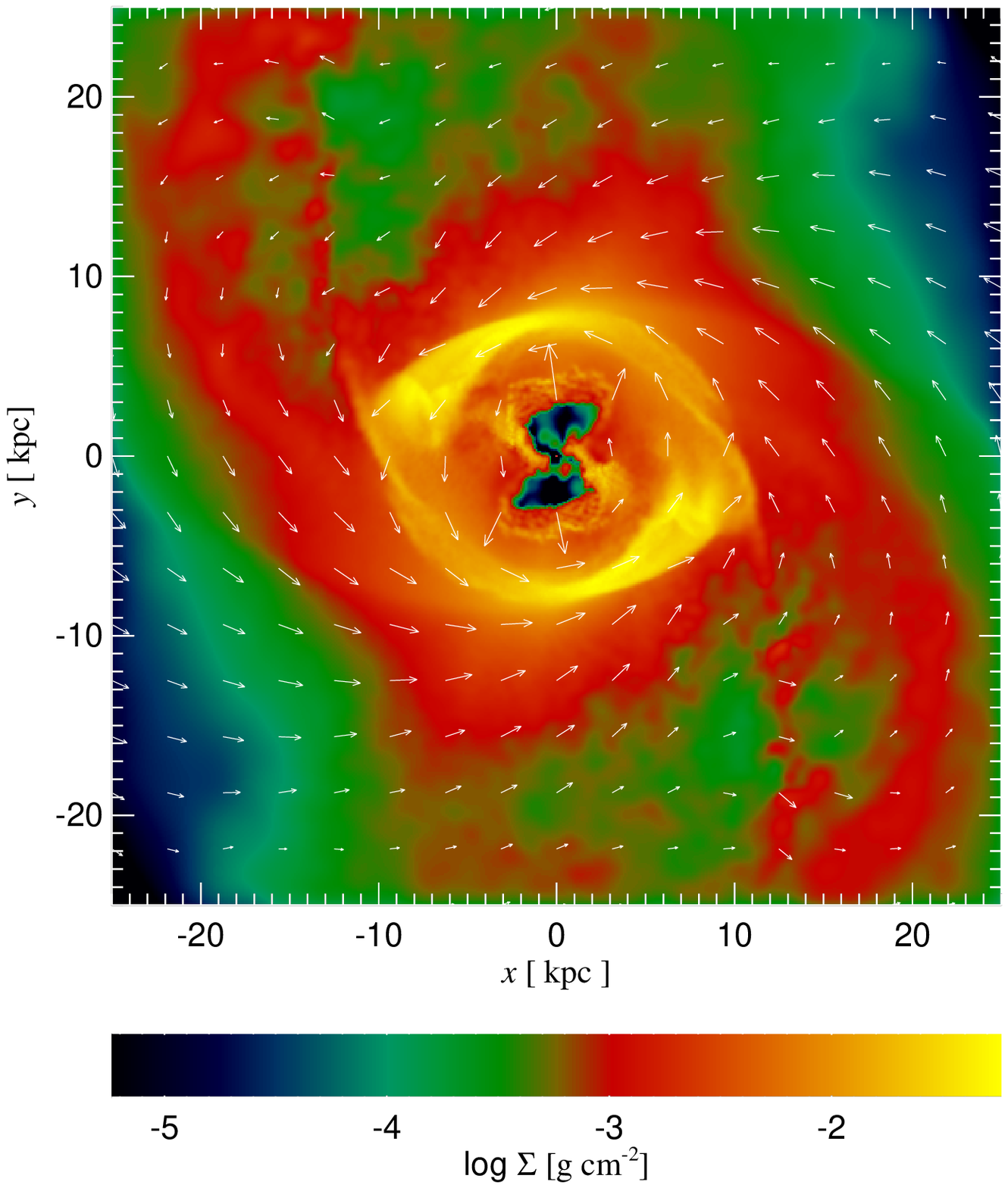,width=0.5\textwidth,angle=0}}
\caption{Left panel -- As in Figure \ref{fig:angle45_fig1} but for time $t=105$
  Myrs; note that the scale has changed. Right panel -- same time but the density
  projection is along the $z$-axis.}
\label{fig:angle45_fig2}

\centerline{\psfig{file=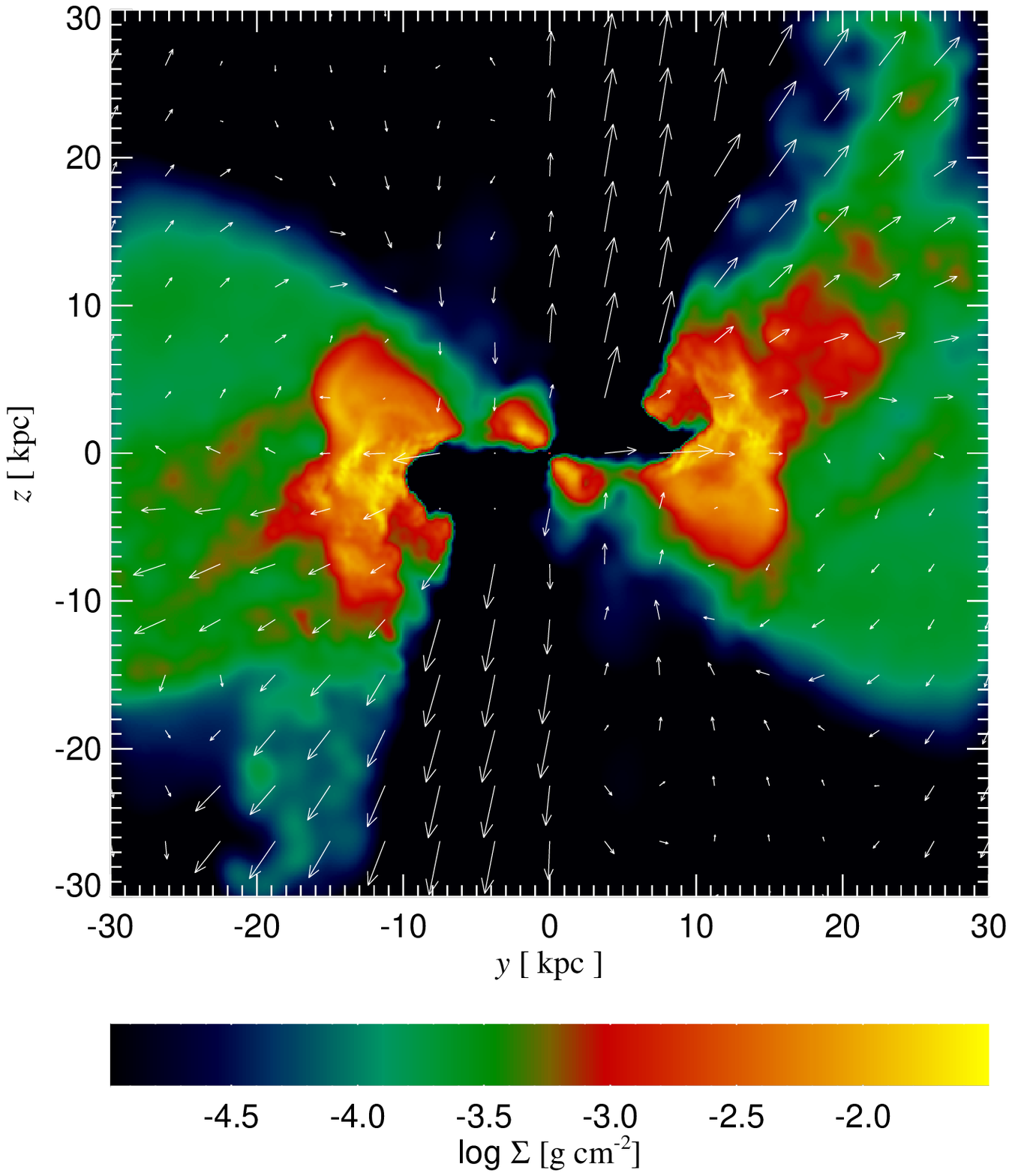,width=0.5\textwidth,angle=0}
\psfig{file=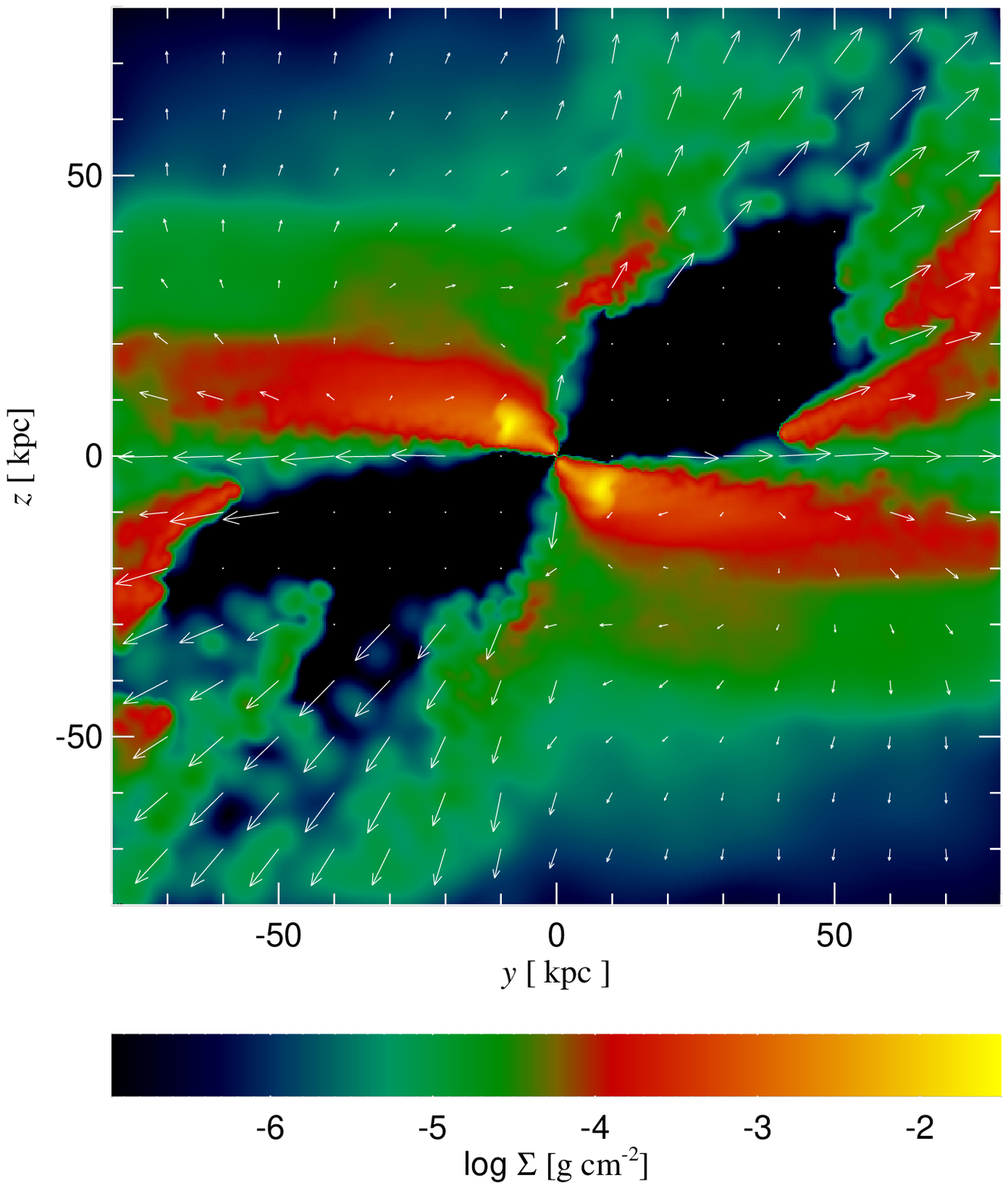,width=0.5\textwidth,angle=0}}
\caption{As in Figure \ref{fig:angle45_fig1} but for later times:
  $t=160$ Myrs and $t=400$ Myrs for the left and the right panels
  respectively. The outflow eventually evacuates the directions along
  which it acts; only the inclined part of the disc that is shielded from the
  feedback survives to late times.}
\label{fig:angle45_fig3}
\end{figure*}

This geometrically complicated interplay of feedback and rotation results in a
very strongly warped rotating structure visible in Figure
\ref{fig:angle45_fig2} at time $t=100$ Myrs. Here we show two projections, one
along the $x$-axis (left panel) and the other along the $z$-axis (right panel)
for clarity. Note the circularising motions of gas at larger radii, settling
into the disc. These regions of the disc are shielded from the black hole
outflow by the inner disc regions. The inner disc regions are highly
disturbed, however, with tangential, outflowing and inflowing motions present
(see the central part of the right panel of Figure \ref{fig:angle45_fig2} ).

Figure \ref{fig:angle45_fig3} shows two later stages in the evolution of the
system. The left panel of the figure is for time $t=160$ Myrs, whereas the
right panel shows $t=400$ Myrs. Apparently the feedback is eventually able to
clear the directions in which it is operating to enforce a strong outflow in
those directions. In short, parts of the disc perpendicular to the outflow do
survive, whereas the less fortunate parts, in the path of the black hole wind,
are blown away by the wind. This occurs on roughly the dynamical time of the
original shell ($R_{\rm out}/\sigma \approx 190$ Myrs). However this is
probably a function of the assumed disc temperature. A cooler and thicker disc
could be more difficult to affect; more simulations of this are needed in the
future.

The resulting ``truce'' between the black hole outflow and the inertia
of the disc is not an easy or natural one. The
disc is warped and is still an evolving structure, with gas orbits
in the inner part being nearly circular while those in the outer parts are
more eccentric. The outflow also has a faster and slower parts, with
the denser regions moving more slowly. Not all parts of the outflow reach
escape velocities before becoming shielded, and some falls back
along directions not exposed to the black hole outflow.

Comparing this simulation to the one with isotropic black hole wind (\S
\ref{sec:feedback_am}), it is apparent that collimated feedback leads the
separation of the gas into disc and outflow regions to occur much later in 
time. Furthermore, this separation is nowhere near as clear cut as in 
\S \ref{sec:feedback_am}. In addition, while no gas was able to reach 
the accretion radius of $R_{\rm acc}=0.5$ kpc in the spherically symmetric 
feedback simulation, the misaligned feedback run resulted in about 1\% of gas 
crossing the inner boundary. This value is still significantly less than the 4\% of gas
accreted in the control simulation with no feedback (\S \ref{sec:nofb}).

\subsection{Outflow with a rotating axis of symmetry}\label{sec:egg_beater}

The final simulation that we present here has a set-up similar to the previous
simulation of misaligned collimated feedback, except in this case the feedback
axis is rotating around the $x$-axis according to:
\begin{equation}
\theta_{\rm out} = \Omega_{\rm out} t\;,
\label{thetar}
\end{equation}
where $\Omega_{\rm out}^{-1} = 30$ Myrs (the simulation in \S
\ref{sec:misalligned} corresponds to a constant $\theta_{\rm out} =
\pi/4$). Thus the outflow axis becomes perpendicular to its original position
in 15 Myrs. Realistically, we might expect changes in the direction of black
hole rotation to occur on even shorter timescales \citep{KingPringle06}, and
so this simulation is probably still less complex than realistic bulges.

\begin{figure*}
\centerline{\psfig{file=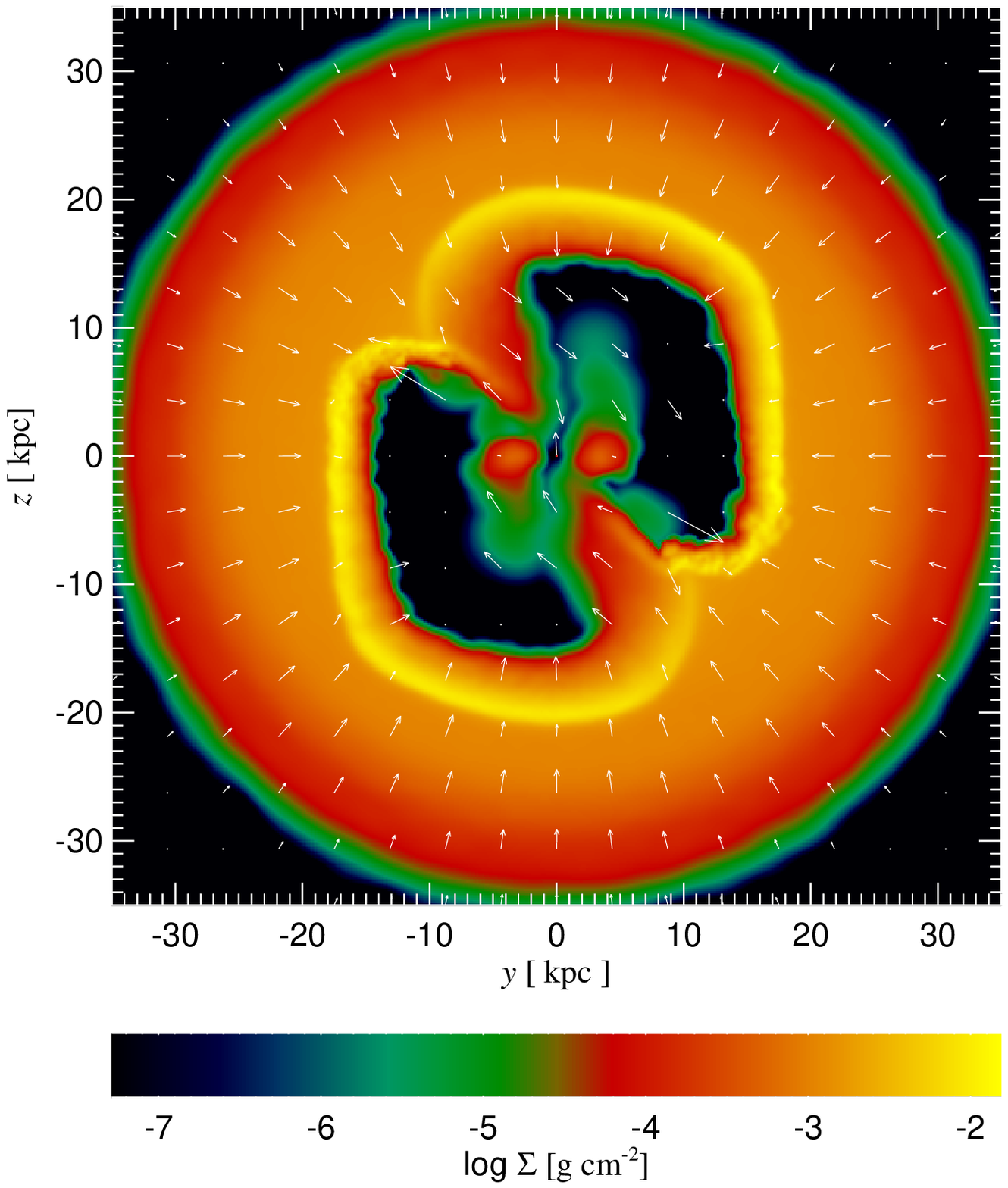,width=0.5\textwidth,angle=0}
\psfig{file=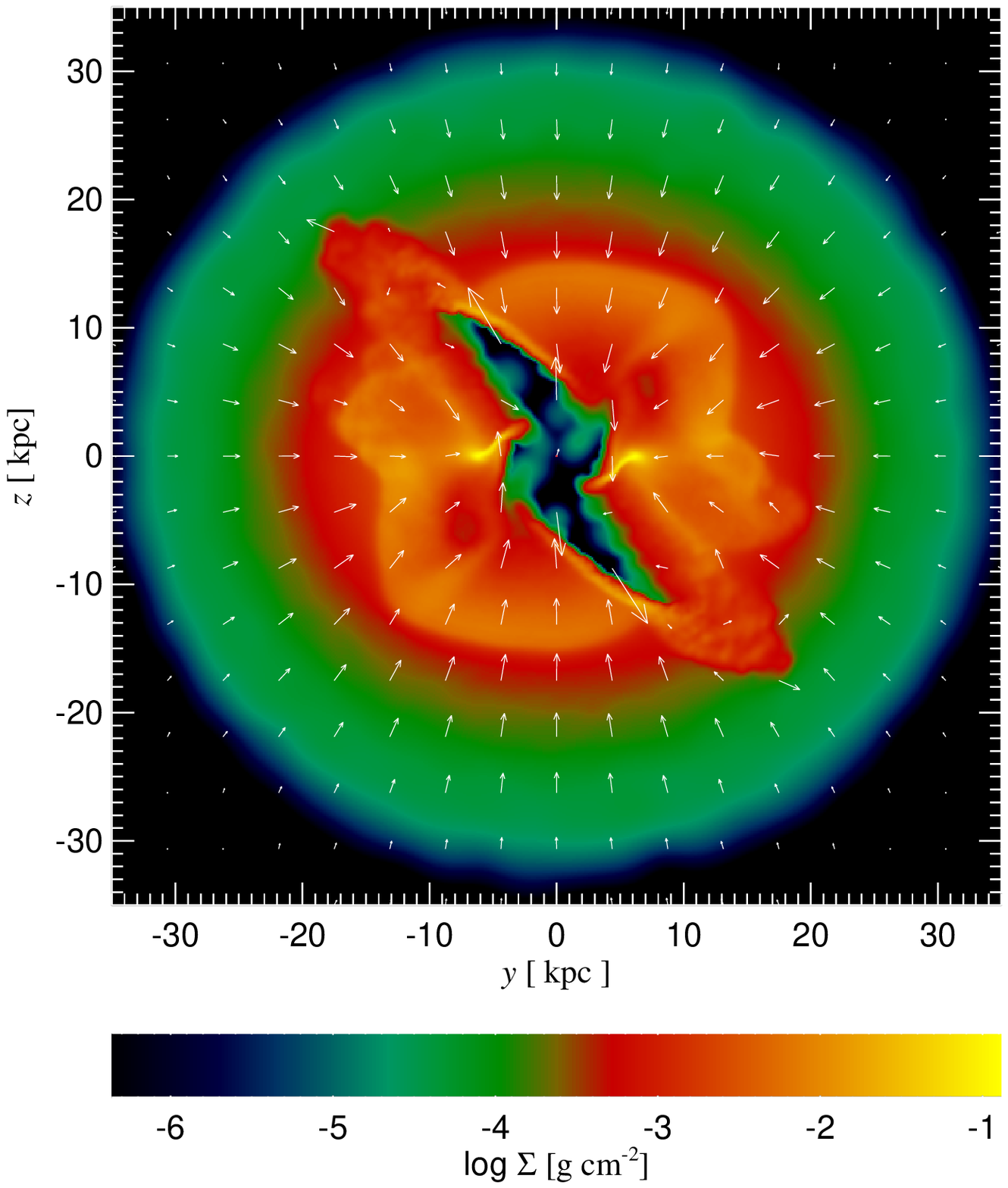,width=0.5\textwidth,angle=0}}
\caption{Simulations with a rotating ``jet'' outflow. Left panel is for
  time $t=47$ Myrs, and right is for $t=83$ Myrs. }
\label{fig:jet_rot_fig1}

\centerline{\psfig{file=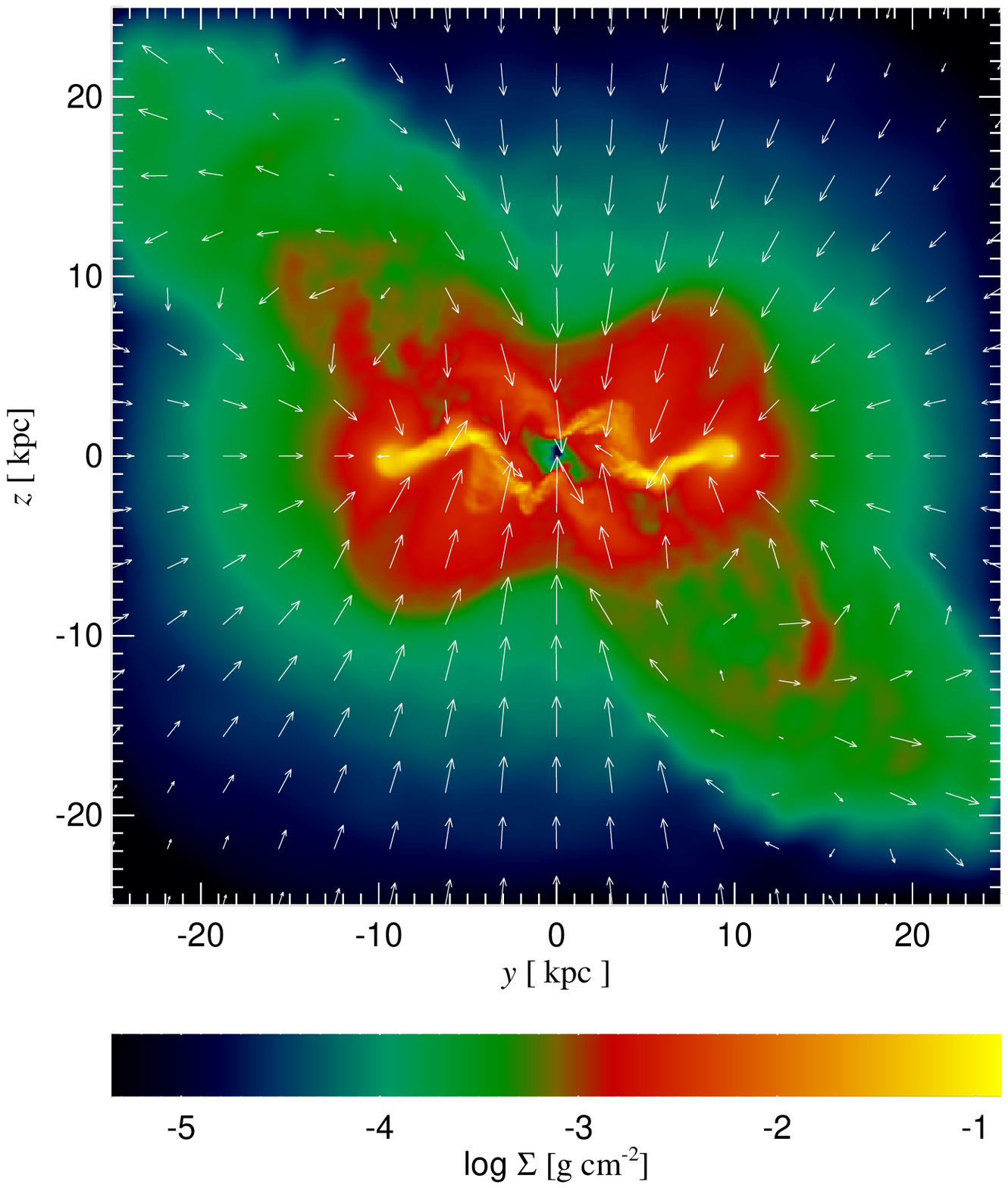,width=0.5\textwidth,angle=0}
\psfig{file=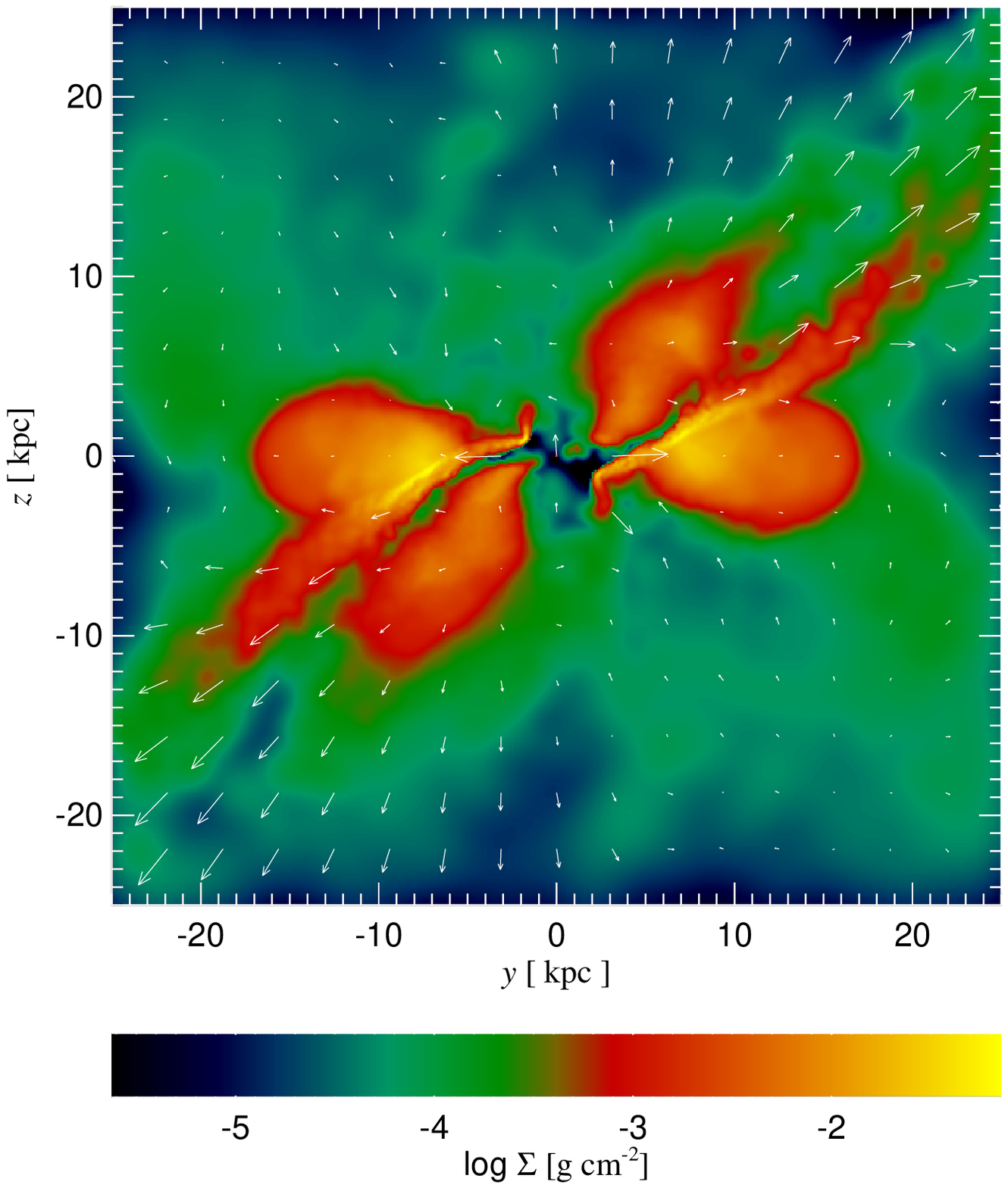,width=0.5\textwidth,angle=0}}
\caption{Same as Figure \ref{fig:jet_rot_fig1} but for later times:
  $t=120$ Myrs and $t=165$ Myrs for the left and the right panels,
  respectively.}
\label{fig:jet_rot_fig2}
\end{figure*}

Figure \ref{fig:jet_rot_fig1} shows the angle-slice projections of the
simulation at times $=47$ (left panel) and $t=83$ (right panel) Myrs. Figure
\ref{fig:jet_rot_fig2} shows the same but at times $=120$ (left panel) and
$t=165$ (right panel). There are certain similarities in gas dynamics between
the present simulation and that with the misaligned collimated outflow (\S
\ref{sec:jet}), but there are also significant differences. One similarity is the
presence of gas inflow along directions not presently exposed to the black hole
outflow, and outflow along some directions. However, there does not appear to
be a settled structure in this case at all. Very strong velocity gradients and
transient shocks occur in the present simulation. Such time dependent shocks
may be efficient drivers of gas turbulence.

\section{Feedback and turbulence}

\begin{figure}
\centerline{\psfig{file=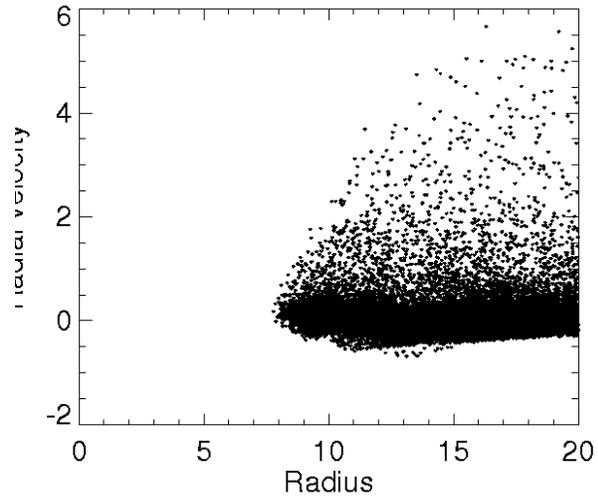,width=0.48\textwidth,angle=0}}
\caption{Radial velocity of SPH particles versus their radial distance
  to the black hole for the isotropic feedback simulation with rotation, see \S
  \ref{sec:feedback_am} and figure \ref{fig:side_L1_later}. The
  dominant feature is rotationally supported disc, and the points
  above it are the SPH particles in the outflow.}
\label{fig:turbulence_fig1}
\end{figure}

Figure \ref{fig:turbulence_fig1} shows the radial velocity structure
for the SPH particles in the isotropic feedback simulation in which
the shell had some angular momentum at a time of $t=236$ Myrs. As we 
remarked in \S \ref{sec:feedback_am}, the feedback and the disc achieve 
a natural division of the available space on the spheres
of influence. Along the directions close to the pole, feedback
dominates. In the midplane, most of the disc particles are shielded by
the inner edge of the disc. The intermediate region is taken up by the
wind driven off from the inner edge of the disc by the feedback. These
features are clearly observable in Figure \ref{fig:turbulence_fig1} in
terms of the radial velocity component.

\begin{figure}
\centerline{\psfig{file=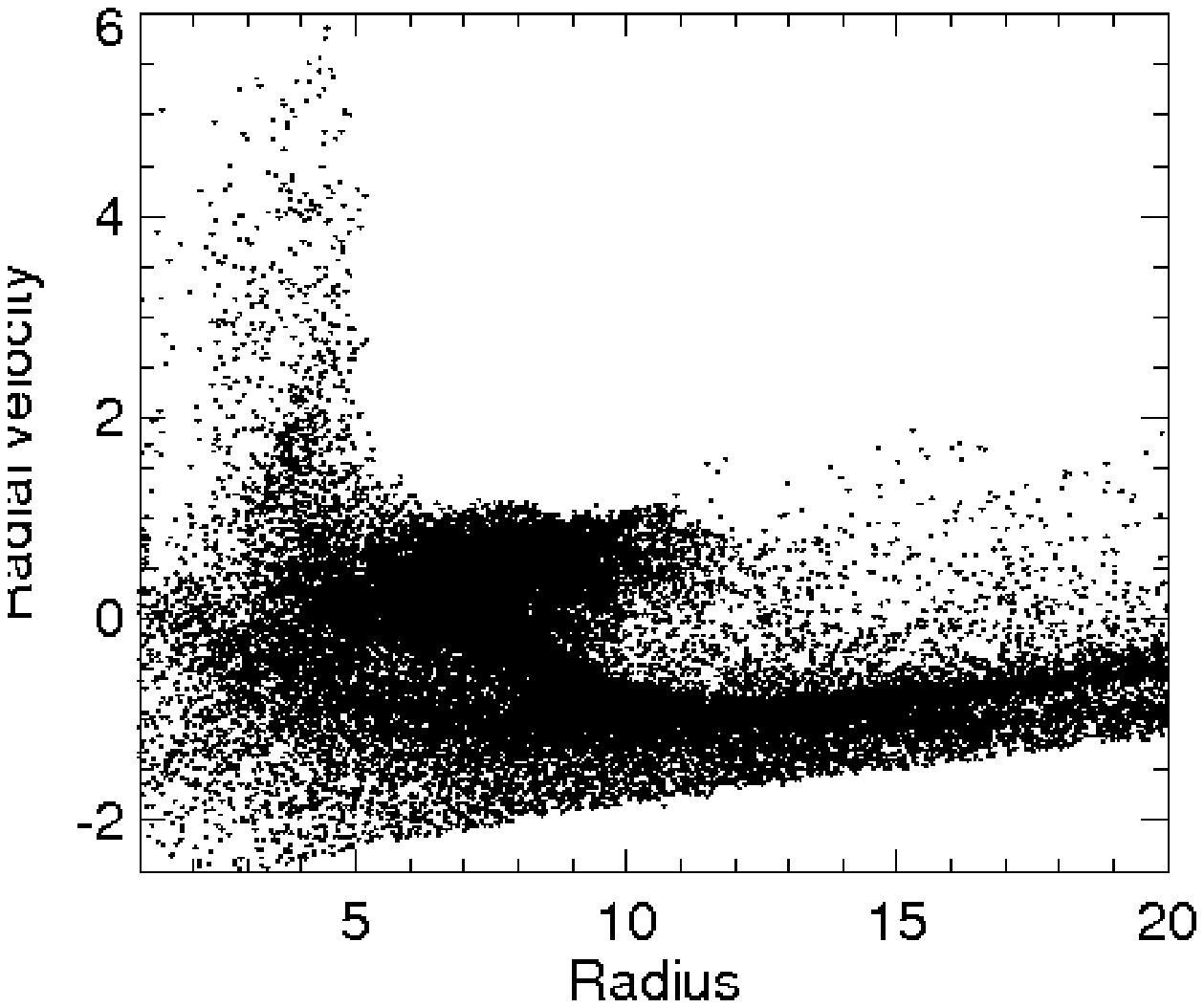,width=0.48\textwidth,angle=0}}
\centerline{\psfig{file=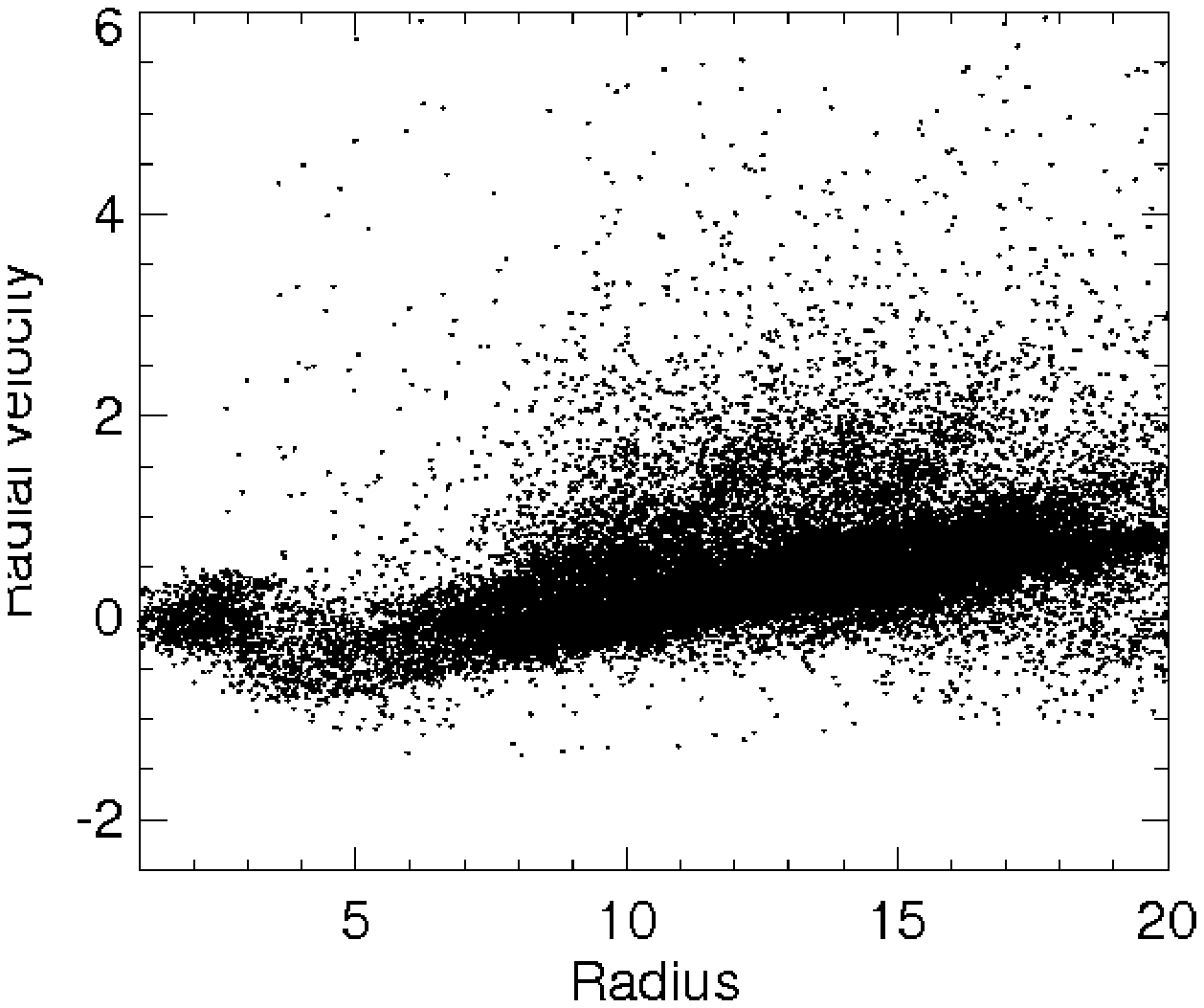,width=0.48\textwidth,angle=0}}
\caption{Same as Figure \ref{fig:turbulence_fig1} but for simulation
  with an inclined collimated outflow (\S \ref{sec:misalligned}) at
  two different times, $t=23$ (upper panel) and $t=59$ (lower panel)
  Myrs. See Figure \ref{fig:angle45_fig1} for the edge-on view of the
  simulation.}
\label{fig:turbulence_fig2}
\end{figure}

We now consider the radial velocity plots for the simulation with a misaligned
(stationary axis) outflow, \S \ref{sec:misalligned}. Figure
\ref{fig:turbulence_fig2} shows that initially the radial velocity
distribution is quite complex, but this settles into the warped disc --
outflow structure fairly quickly already by the time $t=59$ (lower panel) of
the figure. With time this configuration evolves into a progressively more
quasi-stationary state.  

\begin{figure}
\centerline{\psfig{file=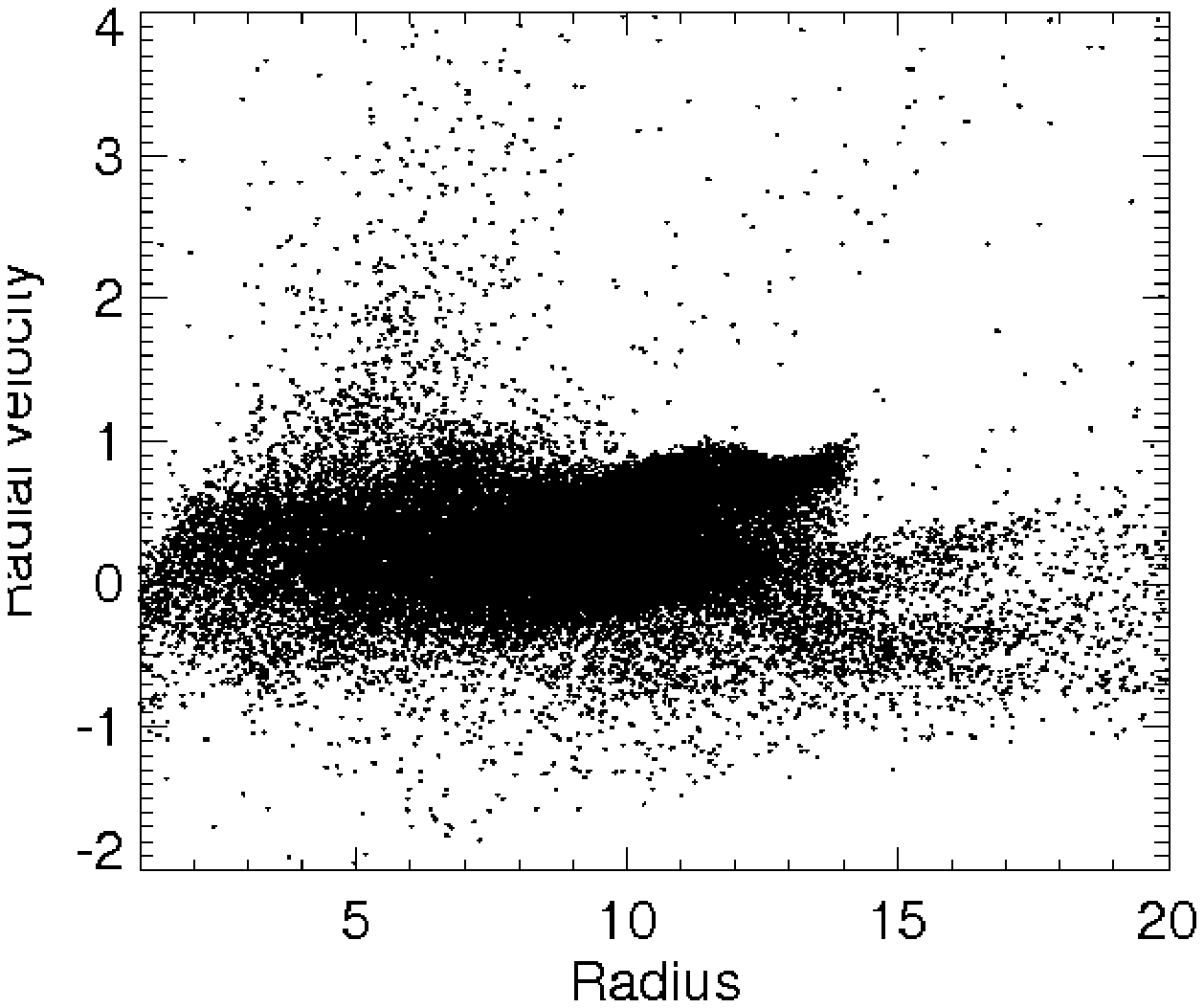,width=0.48\textwidth,angle=0}}
\centerline{\psfig{file=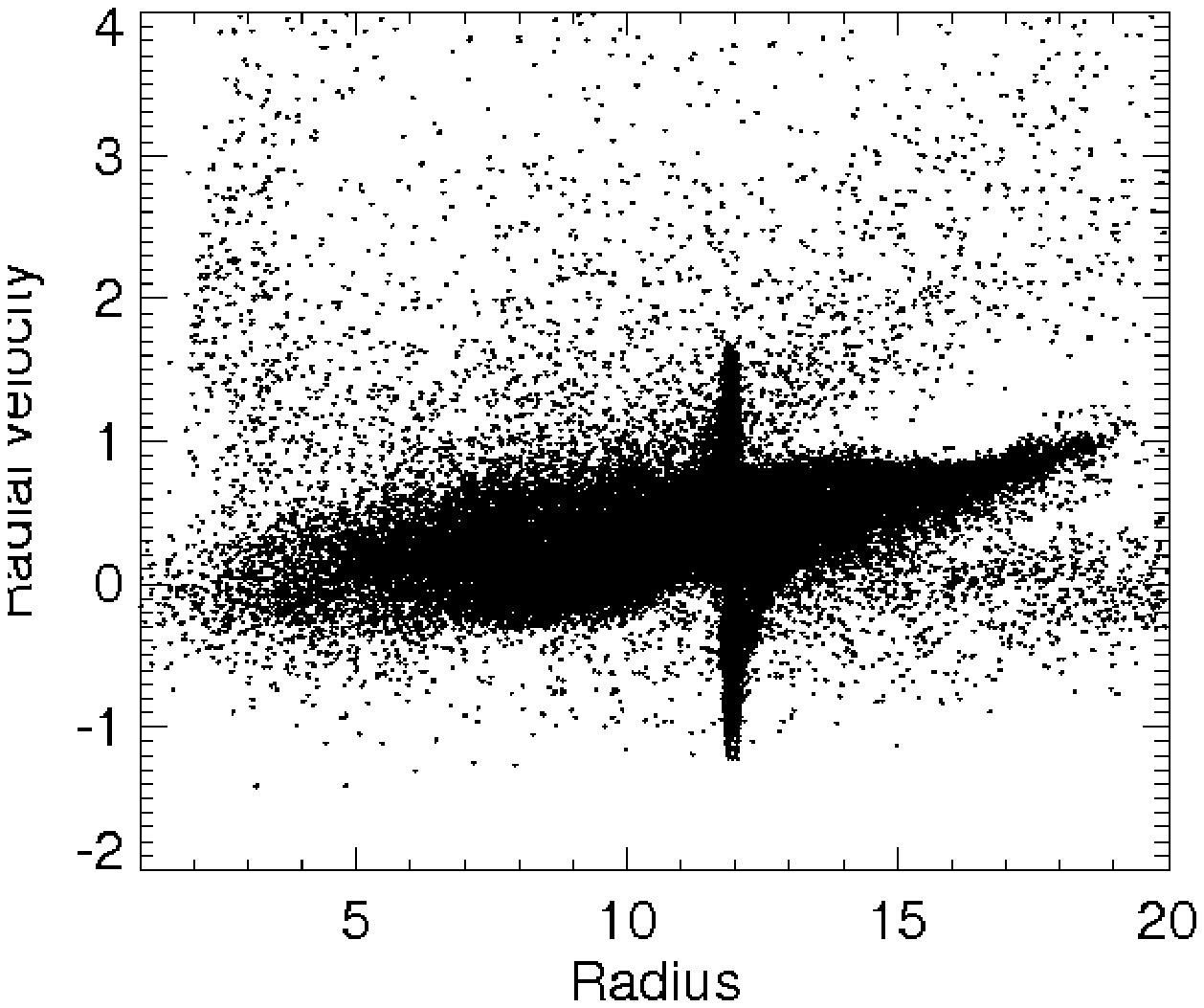,width=0.48\textwidth,angle=0}}
\caption{Same as Figure \ref{fig:turbulence_fig2} but for simulation
  with a rotating collimated outflow (\S \ref{sec:egg_beater}) at
  two different times, $t=141$ (upper panel) and $t=165$ (lower panel)
  Myrs. The vertical feature in the lower panel is a strong transient shock.}
\label{fig:turbulence_fig3}
\end{figure}

In contrast, when the outflow is rotating (Figure \ref{fig:turbulence_fig3}),
no clear cut steady state is reached. The radial velocity pattern keeps
evolving on time scales comparable to the rotation period of the jet at
relatively late times. One observes development of occasional transient shocks
such as the vertical feature in the bottom panel of Figure
\ref{fig:turbulence_fig3}. The SMBH feedback thus provides momentum input into
complicated differential gas motions, probably pumping turbulence in the
bulge.

We believe this is in fact another way in which the SMBH can be affecting its
host galaxy. The current consensus in the field of star formation points to
the key role of turbulence in providing support against gravitational collapse
of gas on large scales while allowing star formation to proceed on smaller
scales in dense filaments \citep[for recent reviews
  see][]{MacLowKlessen04,McKeeOstriker07}. As turbulent motions are not
equivalent to an isotropic pressure support \citep{DobbsEtal05}, the main effect
of the turbulence is to delay or slow down star formation. While future higher
resolution studies are needed to confirm the point, it appears intuitively
clear that AGN-driven turbulence in the bulge can also hinder star formation
in its host bulge. In this case the host is affected by the SMBH feedback in
two ways: not only the gas is being blown away but also the fraction of gas
turned into stars is reduced.

\section{Discussion}\label{sec:discussion}

\subsection{Spherically symmetric models: successful feedback} 

Spherically symmetric models can be considered a success to a certain degree
since earlier analytical work is confirmed by the more detailed calculations
here. First of all, the analytical estimate of the critical $M_\sigma$ mass
obtained by \cite{King03,King05} agrees well with our numerical simulations,
especially for an initially static shell (see \S \ref{sec:static}). If the
shell is already falling in, then the added inwardly directed inertia of the
shell raises the $M_\sigma$ mass by a factor of a few at most (\S
\ref{sec:underweight_fixed} and \S \ref{sec:critical_fixed}).

The complications arising from a finite system size were recently pointed out
by \cite{NayakshinEtal09b} who noted that dynamical time in the bulge,
$R_b/\sigma$, where $R_b$ is the scale radius of the bulge, plays a key role
in what happens. Black holes growing at the Eddington rate, $\dot M_{\rm
  Edd}$, $e$-fold their mass on the Salpeter time, $t_{\rm S} = M_{\rm
  BH}/\dot M_{\rm Edd} \approx 4.5\times 10^7$. Observationally,
\begin{equation}
t_{\rm dyn} = \frac{R_b}{\sigma} = 17 \hbox{Myrs}
\left[\frac{\sigma}{150 \hbox{ km s}^{-1}}\right]^{2.1}\;.
\label{tdyn}
\end{equation}
Thus ``large'' bulges, $\sigma \simgt 200$ km s$^{-1}$, have dynamical times
comparable or larger than $t_{\rm S}$. Under the assumption that the duration
of the black hole feeding event is comparable to the dynamic time of the bulge
then, these ``large'' bulges allow sufficient time for their SMBH to grow
significantly. These bulges would then have ``fully grown'' black holes
saturated at their $M_{\sigma}$ mass. On the other hand, lower mass bulges
with velocity dispersions of $\sigma \simlt 100$ km s$^{-1}$ would evolve too
rapidly for their SMBHs to ``catch up''. Therefore feedback from these SMBH
would be overpowered by the inflow of gas from the bulge. The central regions
of these lower mass bulges could be sites of copious star formation, leading
to the birth of nuclear star clusters. The latter then reach their $M_{\sigma
  *}$ masses \citep[see][]{NayakshinEtal09b}, replacing the SMBH black holes
as the dominant objects in the bulges.

Our simulations (\S \ref{sec:eddington}) confirm this effect. The numerical
experiment in \S \ref{sec:large} has shown that black hole immersed in a 40
kpc bulge is able to grow enough to expel the gaseous shell when started from
initial mass of $M_{\rm bh} = 10^8 \msun$ (Fig \ref{fig:M0.01_R40_edd}). At
the same time, a black hole that is initially twice as massive in a 10 kpc
bulge is unable to expel the shell (\S \ref{sec:small},
Fig. \ref{fig:M0.02_R10_edd}) that is less massive than that in \S
\ref{sec:large}.

One complication not considered by analytical models yet is that if the mass
fraction of gas is significant, i.e. larger than the universal cosmological
fraction of $f_g \approx 0.16$, then the self-gravity of the gas can dominate the
potential. The weight argument of course applies to self-gravitating shells as
well, and they too can be unbound by a sufficiently strong black hole wind
(e.g., see Figure \ref{fig:M0.01_R40_edd}). However in this case our models
are inconclusive as they do not include star formation. When gas is blown
away, the velocity dispersion of the self-gravitating shell should decrease. A
detailed calculation including star formation is needed to establish whether
or not the black hole in the remaining bulge follows the $M-\sigma$ relation.

\subsection{Spherically symmetric models: inconsistent SMBH feeding}

There are at least two major worries about the general usefulness of
spherically symmetric models. Large scale cosmological simulations usually
produce very complicated gas density and velocity flows onto the centres of
dark matter halos \citep[e.g.,][]{KeresEtal05} and so it is not clear
whether or not such spherically symmetric flows ever take place.

There is also a conceptual difficulty with the spherically symmetric
models. In a self-consistent model, where black hole feedback is linked to the
accretion of gas onto the hole, one cannot have accretion and outflow at the
same time. Yet both are needed. We need accretion to grow and power the
SMBH, but we also need outflow to eventually curtail the SMBH growth and expel the
gas from the galaxy. 

Having accretion before, e.g., while the SMBH has not reached the $M_\sigma$
mass, and outflow later, after it has done so, is not a comfortable
option. Indeed, to change inflow to outflow for most of the gas in a bulge,
we require momentum input of 
\begin{equation}
P \sim M_{\rm gas} v_{\rm esc} \sim M_{\rm bulge} \sigma\;,
\end{equation}
where $M_{\rm gas}$ is the mass of the gas which we assumed to be of order the
bulge mass, $M_{\rm bulge}$, and the escape velocity $v_{\rm esc} \sim
\sigma$. The minimum amount of mass that the black hole needs to accrete to
produce this much momentum outflow is
\begin{equation}
\Delta M_{\rm min} = { P \over \epsilon c}\;.
\end{equation}
For fiducial numbers $\sigma = 150$ km/sec and $\epsilon = 0.1$, we have 
\begin{equation}
\Delta M_{\rm min} = 0.005 M_{\rm bulge} \;,
\end{equation}
which is comparable to the observed $M_{\rm bh}$ masses for most bulges
\citep{Haering04}. Now, if this gaseous mass is accumulated somewhere near the
SMBH, e.g., in a disc that is immune to the feedback so that it can power the
SMBH, then the question is why could this mass not be much higher or much
lower?

In other words, if SMBH feeding is local to a small scale region (for example,
inside the SMBH influence radius, which is typically between a pc and a few
tens of pc), where enough material is stored to feed SMBH while most gas is
driven away, the casual link between feedback and accretion is broken, and
there does not appear a reason to have an $M_{\rm bh} - \sigma$ relation. Thus
spherically symmetric models, while providing excellent fits to the observed
relations, do not naturally explain how the SMBH gets its fuel, or how it knows
to stop growing once the many kpc scale shell starts to be blown away.

\subsection{Non-spherical models: clear mode for feeding but not feedback}\label{sec:am_role}

The second type of simulations that we considered here do not have spherical
symmetry for one or more reasons. SMBH feeding in these models can be achieved
(1) by features denser than the mean, i.e., discs or filaments, or (2) through
regions not exposed to feedback at all. In the former case, if the
SMBH-feeding gas has a column density (as seen from the black hole) much
higher than the mean for the bulge, then it is only weakly affected by the
momentum feedback. In both of these cases there can be inflow to feed the SMBH
and outflow to expel most of the gas in the bulge simultaneously. However, if 
these dense features feed the SMBH despite it producing feedback at the
maximum rate, then it is not clear why there should be a feedback mediated 
SMBH -- galaxy link.

This problem is already apparent in the simplest non-spherical case -- the
simulation of a rotating initially spherical shell of gas (\S
\ref{sec:feedback_am}). In that simulation gas separates out in two well
defined regions -- the polar region where black hole outflow dominates and the
disc midplane where gas is shielded from the black hole outflow by the inner
edge of the disc. The edge of the disc is slowly ``evaporating'' into an
outflow. The rest of the disc is hardly affected, and the orbits there are
nearly circular.

We have also set a numerical experiment in \S \ref{sec:misalligned} where
feedback outflow is confined to a broad cone misaligned with the rotation
axis of the shell. The shell initial condition was identical to that explored
in \S \ref{sec:feedback_am}. We found, not surprisingly, that the resulting
pattern of gas dynamics is even more complex than that of an isotropic
feedback with a rotating shell. The inflow persists along directions in which
black hole winds are not emitted, and the outflow prevails inside the feedback
cone. The collimated rotating black hole outflow (\S \ref{sec:egg_beater}) is
even more complex geometrically, with black hole feedback probably driving
turbulence into bulge gas.

\subsection{Shortcomings and comparison to other numerical work}

There is a substantial body of work on numerical simulations of SMBH feedback
in cosmological simulations, e.g.,
\cite{SpringelEtal05,DiMatteo05,SijackiEtal07}. This paper complements this
work. While addressing similar issues, the scales involved in studying SMBH
feedback in a cosmological context are very different. Cosmological
simulations have clear strengths -- they naturally account for hierarchical
build-up of galaxies assuming realistic initial conditions and they include
recipes for physical processes that we expect to be important for galaxy
formation such as cooling, star formation and feedback. Therefore they can
provide a powerful tool with which to explore SMBH feedback. In comparison,
our simulations are more modest in terms of the number of physical processes
studied, making use of idealised initial conditions in static analytical
potentials.

On the other hand, the all-inclusive nature of cosmological
simulations mean that ``individual'' processes, such as black hole
accretion and feedback, cannot be modelled in sufficient physical detail and
require sub-grid physical models to be assumed. As far as we are aware,
all the current cosmological simulations rely on \cite{Bondi44,Bondi52}
formulation for accretion of gas. The latter formulation is physically
inconsistent if accreting gas has angular momentum as gas simply cannot reach
the SMBH due to the centrifugal barrier. Furthermore, at least the ``quasar''
mode of feedback is done in a very sketchy manner, with a fixed fraction of
energy released by the quasar passed to the SMBH neighbour particles. The right
fraction is found a posteriori to fit the observed $M_{\rm bh} - \sigma$
relation and other observational constraints. 

Therefore, modest ``small scale'' simulations such as ours, especially when extended
to include more physics and to reach in as close as possible to the SMBH may
offer another important route to solving the mystery of SMBH and galaxy
growth. Ideally, the small scale and cosmological simulations should be merged
to some degree in the future to allow as self-consistent a study of SMBH
feedback as possible.

\section{Conclusions}

Our main conclusions are
\begin{itemize}

\item Predictions of spherically symmetric models with black hole feedback
  tied to Eddington limit luminosity as in the models of \cite{King03,King05}
  are confirmed numerically. Self-gravity of the gas complicates evolution of
  the system, and a self-consistent treatment of star formation and its
  feedback is necessary for further progress.

\item As suggested earlier based on analytical arguments, it is the dynamical time in
  the bulge, $R_{\rm b}/\sigma$, that determines whether or not the SMBH can reach its
  limiting $M_\sigma$ value. Central regions of smaller bulges, where the
  dynamical time is observed to be shorter than the Salpeter time, could be
  smothered with infalling gas despite the SMBH feedback. This process may be
  the origin of the nuclear star clusters in ``smaller'' galaxies.

\item A net angular momentum in the shell is essential in determining
  the fate of the shell and the SMBH feeding. There is no well defined
  $M_\sigma$ mass in that case since the momentum thrust required is
  different in different directions. Gas near the symmetry axis is
  blown out easier than in the spherically symmetric case, whereas gas
  settled into a disc requires $\sim R/H$ more thrust than in the
  latter case.

\item If the SMBH is fed through a several kpc scale disc, the SMBH mass would
  have to be $\sim R/H$ larger than in the spherical case to expel the
  feedback-resistant self-shielding disc. This directly contradicts the recent
  observations of \cite{Hu09} that show that pseudo-bulges have lighter SMBHs
  than their classical counter-parts. This may imply that black holes are not
  fed by large (kpc or larger) discs but rather by flows with a small specific
  angular momentum.

\item We also noted that SMBH outflows in a realistic, i.e., aspherical,
  situation may pump turbulence (differential motions) in the
  bulge. Turbulence is known to hinder star formation. Therefore we believe
  that AGN feedback not only expels the gas from the galaxy but it may also
  reduce the amount of mass turned into stars during bulge formation.

\end{itemize}

\section{Acknowledgments}

The authors thank Andrew King for illuminating discussions and comments on the
draft of the paper. Volker Springel is thanked for help with Gadget and
comments on an early draft of the paper. Theoretical Astrophysics research at
the University of Leicester is supported by a STFC Rolling grant.


\label{lastpage}

\end{document}